\documentclass[12p, a4paper, twocolumn]{article}

\usepackage{amsmath,amssymb,amsfonts} 
\usepackage{hyperref}
\hypersetup{colorlinks=true, urlcolor=blue, linkcolor=blue, citecolor=blue}
\usepackage{graphicx} 
\usepackage{cite}

\usepackage{xr} 
\begin{document}
\date{}

\title{Hierarchical friction memory leads to subdiffusive configurational dynamics of fast-folding proteins}
\author{Anton Klimek, Benjamin A. Dalton, Lucas Tepper, Roland R. Netz\\
\\
\textit{Fachbereich Physik, Freie Universit{\"a}t Berlin, Arnimalle 14, 14195 Berlin, Germany}}





\newcommand{\abstractText}{
    Proteins often exhibit subdiffusive configurational dynamics.
    The origins of this subdiffusion are still unresolved.
	We investigate the impact of non-Markovian friction and the free energy landscape on the dynamics of fast-folding proteins in terms of the mean squared displacement (MSD) and the mean first-passage-time (MFPT) of the folding reaction coordinate.
	We find the friction memory kernel from published molecular dynamics (MD) simulations to be well-described by a hierarchical multi-exponential function, which gives rise to subdiffusion in the MSD over a finite range of time.
	We show that friction memory effects in fast-folding proteins dominate the scaling behavior of the MSD compared to effects due to the folding free energy landscape.
	As a consequence, Markovian models are insufficient for capturing the folding dynamics, as quantified by the MSD and the MFPT, even when including coordinate-dependent friction.
	Our results demonstrate the importance of memory effects in protein folding and conformational dynamics and explicitly show that subdiffusion in fast-folding protein dynamics originates from memory effects, not from the free energy landscape and not from coordinate-dependent friction.
}

	\twocolumn[
	\begin{@twocolumnfalse}
		\maketitle
		\begin{abstract}
			\abstractText
			\newline
			\newline
		\end{abstract}
	\end{@twocolumnfalse}
	]
	
	\section{Introduction}
	Almost every biological process relies on the functioning of proteins, which is strongly dependent on their specific three-dimensional arrangement.
    Whereas recent tools such as alpha fold allow fast and accurate predictions of the folded structure of a protein for a given amino-acid sequence \cite{jumper2021highly}, they do not offer any information about the dynamics of the folding that leads to that conformation.
    However, dynamical information is crucial for many applications, such as the development of interventions of misfolding \cite{sweeney2017protein}.
	Since a fully atomistic description of a protein involves many degrees of freedom, the structure is often projected on low-dimensional reaction coordinates (RC) \cite{fukui_formulation_1970,best_reaction_2005}, or coarse-grained into subunits \cite{kmiecik2016coarse,heath2007coarse,monago2025unraveling}.
	The choice of a suitable RC is highly nontrivial.
	Usually, one assumes the RC to be Markovian \cite{krivov_reaction_2013}, i.e. to exhibit no dependence on past dynamics but only on the present state, in which case a description of the dynamics in terms of static friction and the free energy landscape of the RC is possible \cite{kramers_brownian_1940}.
    Alternatively, one can use Markov state models to identify protein folding pathways \cite{shukla2015markov,malmstrom2014application}.
	However, it has recently been shown that, in the case of protein folding, the dynamics of many RCs exhibit significant non-Markovian behavior \cite{ayaz_non-markovian_2021,dalton_fast_2023}. 
    This suggests that finite timescale intramolecular and solvent relaxation processes are significant and cannot be neglected.
    In addition, experimentally only certain RCs are measurable, which makes it practically impossible to observe a Markovian RC.
	Thus, there have been numerous studies that focus on non-Markovian effects in protein conformational dynamics \cite{min2005observation,ayaz_non-markovian_2021,kou_generalized_2004,plotkin_non-markovian_1998,berezhkovskii_single-molecule_2018,dalton_fast_2023,vollmar_model-free_2024,dalton2024ph}.
    These studies rely on the generalized Langevin equation (GLE), an exact equation derived from the underlying many-body Hamiltonian \cite{mori_transport_1965,ayaz_generalized_2022,dalton2024memory}.
	For a given RC $q(t)$, the GLE reads
	\begin{equation}
		\label{eq_gle_wom}
		m \ddot{q}(t)=-\nabla U(q(t)) -\int_{0}^{t} \Gamma(t-t')\dot{q}(t') dt' + F_R(t) \,,
	\end{equation}
	where $-\nabla U(q)$ is the force due to the free energy landscape $U(q)$, $\dot{q}(t)$ denotes the velocity and $\ddot{q}(t)$ the acceleration of the RC, and $\Gamma(t)$ is the friction memory kernel, which incorporates the non-Markovian effects.
    The friction kernel $\Gamma(t)$ and the random force $F_R(t)$, as well as the free energy $U(q)$ originate from the projection of the many-body dynamics on a single RC.
	In an equilibrium scenario, the mass, which is assumed here to be independent of $q$, is defined by an analog the equipartition theorem as $m=k_{\rm{B}}T/\langle \dot{q}^2 \rangle$, and the random force $F_R(t)$ fulfills the relation 
	\begin{equation}
		\label{eq_fdt}
		\langle F_R(t) F_R(0) \rangle = k_{\rm{B}}T \Gamma (t)\,.
	\end{equation}
	Eq.~\eqref{eq_fdt} is exact for the Mori GLE \cite{mori_transport_1965}, for which $U(q)$ is harmonic; however, for non-harmonic $U(q)$, Eq.~\eqref{eq_fdt} is approximate \cite{vroylandt2022derivation,ayaz_generalized_2022}.
    The validity of these approximations on $m$ and $F_R$ is confirmed later by the accurate agreement of GLE predictions with MD simulations.
    
	A key measure to describe the dynamics of the RC is the mean squared displacement (MSD), defined as $C_{\textrm{MSD}}(t) = \langle (q(0) - q(t))^2 \rangle$, where $\langle \cdot \rangle$ denotes the average of the ensemble.
    The MSD is often written as $C_{\rm{MSD}}(t)\propto t^{\alpha(t)}$ with a time-dependent exponent $\alpha(t)$.
    In the free case with $U(q)=0$ and with instantaneous friction $\Gamma(t)\propto \delta(t)$, i.e. the Markovian case, the MSD exhibits a ballistic regime defined by $\alpha=2$ at short times followed by a transition to Brownian diffusion with $\alpha=1$ at long times.
	Additional features in the MSD, such as oscillations or subdiffusive scaling (the latter being characterized by $\alpha(t)<1$), can in principle arise either from non-Markovian friction $\Gamma(t)$, from features in the free energy landscape $U(q)$, from coordinate-dependent friction \cite{best2010coordinate, hinczewski_how_2010}, or from a combination of these effects.
    The interplay of these effects and how they lead to subdiffusion in protein conformational dynamics is still debated.

	Theories of subdiffusion often postulate power-law memory $\Gamma(t)\propto t^{-\alpha}$, motivated by the idea that the systems of interest exhibit self-similar structures and hence display fractal patterns in the dynamics on all time scales \cite{kou_generalized_2004, metzler_random_2000, kappler_cyclization_2019}.
    However, real-world physical systems are naturally limited by a smallest and largest temporal and spatial scale, which necessarily limits the self-similarity \cite{noauthor_advances_2012}.
    This is certainly the case for proteins, which are atomic in composition at their shortest scales and fold into structures that are typically of the order of nanometers in size.
    Interestingly, proteins are known to exhibit a wide range of relevant time scales, spanning from solvent interactions at the picosecond scale, up to microseconds and even seconds.
    However, in simulations as well as in experiments this wide range of time scales is often not completely captured, which illustrates why protein folding is often described in terms of fractals and power-law models over all available time scales of the data.
	Previous theoretical works revealed that power-law memory can be accurately described by sums of very few exponential contributions and that such multi-exponential memory contributions produce subdiffusion in the MSD over many orders of magnitude in time \cite{ayaz_non-markovian_2021,noauthor_advances_2012,klimek2024_theo}.

    The mean first-passage-time (MFPT) is another key measure that links the conformational dynamics of proteins to their folding and allows us to evaluate folding times.
    It is defined as the mean time to reach the final RC position $q_F$ for the first time, from a starting position $q_S$.
    The MFPT from the unfolded to the folded state is called the folding time and vice versa for the unfolding time.
	Folding and unfolding times of proteins are measurable in experiments, they are strongly influenced by non-Markovian effects \cite{dalton_fast_2023} and are highly relevant for their biological functioning \cite{hartl2017protein}.
	
	In this paper, we extract the friction memory $\Gamma(t)$ of several fast-folding proteins from published MD simulation trajectories \cite{ayaz_non-markovian_2021,lindorff-larsen_how_2011}.
    The proteins ($\lambda$-repressor, $\alpha_3D$, protein-G and the homopeptide Alanine9), chosen to cover a wide range of free energy barrier heights, all exhibit restricted power-law friction memory for well-established one-dimensional RCs.
    The friction kernels are all well described by very few exponential contributions.
    This multi-exponential memory effectively produces power-law behavior for a finite range of time in the MSD.
	We find that the subdiffusive regime in the MSD is well captured by the GLE with multi-exponential memory but cannot be reproduced by Markovian models.
    Moreover, we show that the GLE describes the MFPTs well for both, folding and unfolding, unlike Markovian models, even if they include coordinate-dependent friction.
	
	The extraction and fit of the friction memory kernels allow us to accurately predict the subdiffusive MSD regime using analytic theory \cite{klimek2024_theo}, where the memory is multi-exponential and individual memory components are hierarchically ordered.
	In conjunction with GLE simulation, this makes it possible for us to disentangle the influences of the free energy and of the friction on the conformational dynamics of proteins.
	The observed hierarchical relationship between friction memory components further enables us to estimate memory contributions below the temporal resolution of the data, the inclusion of which leads to an even better description of the MSD.
	Our results highlight the importance of memory effects in protein conformational and folding dynamics and specifically show that memory dominates the dynamics of fast-folding proteins compared to effects due to the free energy landscape and compared to Markovian coordinate-dependent friction effects.
    We find that fast-folding proteins can be described by hierarchically ordered friction-memory components, leading to subdiffusive conformational dynamics, which reconciles the ideas of hierarchical and fractal protein folding theories \cite{baldwin_is_1999,mori_molecular_2016,sangha_proteins_2009,satija_transition_2017,hu_dynamics_2016}.

	\begin{figure*}
		\vspace{-42mm}
		\includegraphics[width=\textwidth]{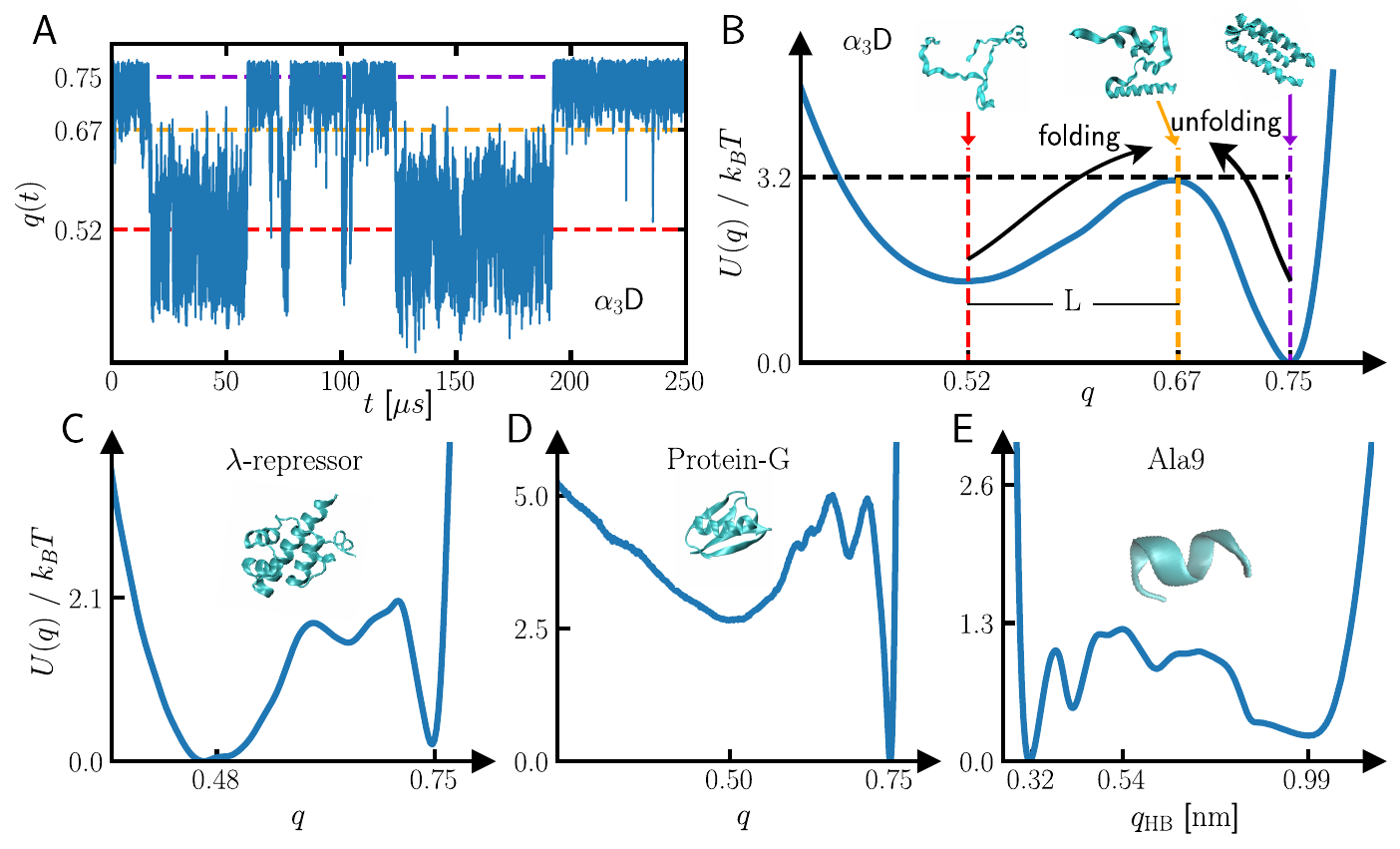}
		\caption{(A) Trajectory of the fraction of native contacts RC $q(t)$ from atomistic MD simulations of the $\alpha_3$D protein \cite{lindorff-larsen_how_2011}.
			(B) The free energy profile $U(q)$ extracted from the $q(t)$ trajectory for the $\alpha_3$D protein, evaluated using Eq.~\eqref{eq_boltzmann_inverted}.
			Colored vertical dashed lines denote the minima and the maximum of $U(q)$, with exemplary snapshots of the folded, unfolded, and barrier-state conformations above the respective RC positions.
			$L$ denotes the distance between the unfolded state and the barrier state.
			The extracted free energy profiles for (C) $\lambda$-repressor, (D) protein-G and (E) Ala9, including snapshots of the folded-state conformations, where $q_{\rm{HB}}$ is the $\rm{HB}_4$ RC for Ala9 \cite{ayaz_non-markovian_2021}, which has the value $q_{\rm{HB}}=0.32\,\rm{nm}$ in the folded state.}
		\label{fig1}
	\end{figure*}
    
	\section{Modeling \& Results}
	The folding of proteins without energy consumption is an equilibrium process \cite{englander2014nature,thirumalai2001chaperonin}, thus the GLE (Eq.~\eqref{eq_gle_wom}) together with Eq.~\eqref{eq_fdt} are appropriate to describe the dynamics and the RC is distributed according to the Boltzmann distribution $p(q)\propto e^{-U(q)/k_{\rm{B}}T}$.
	Inversely, the free energy is extracted from trajectories via
	\begin{equation}
		\label{eq_boltzmann_inverted}
		U(q) = -k_{\rm{B}}T \ln(p(q))\,.
	\end{equation}
	Fig.~\ref{fig1}A shows a $250\,\mu\rm{s}$ trajectory segment for the fraction of native contacts RC taken from an extensive MD simulation of the fast-folding protein $\alpha_3D$, which exhibits transitions between the folded and unfolded states every few microseconds \cite{lindorff-larsen_how_2011}.
	Fig.~\ref{fig1}B depicts the free energy profile $U(q)$ extracted from such trajectories by Eq.~\eqref{eq_boltzmann_inverted}, which exhibits two local minima corresponding to the folded and unfolded state, illustrated by exemplary snapshots of the protein structure at the two minima and the barrier-top.
	
    The free energy landscapes and native state conformations of all other proteins in this study are shown in Figs.~\ref{fig1}C-E, which exhibit free energy barriers with heights $U_0$ between $1\,k_{\rm{B}}T$ and $5\,k_{\rm{B}}T$.
	The proteins are $\lambda$-repressor, a transcription inhibitor in bacteriophages \cite{stayrook2008crystal}, protein-G, present in the cell walls of bacteria and involved in antibody binding \cite{bjorck1984purification}, $\alpha_3D$, a designed $\alpha$-helical fast-folding protein \cite{zhu2003ultrafast}, and Ala9, a nine residue homo-peptide that forms a single $\alpha$-helix \cite{kuczera2024helix}.    
	The temporal resolution of the MD simulation data of the fast-folding proteins \cite{lindorff-larsen_how_2011} is $\Delta=0.2\,\rm{ns}$, the resolution of the Ala9 simulation data is \cite{ayaz_non-markovian_2021} $\Delta=1\,\rm{fs}$.
	It was shown that the temporal resolution of the analyzed MD data leads to valid friction kernels $\Gamma(t)$, where only the information on times below the discretization is obscured \cite{dalton_fast_2023, tepper2024accurate}.
	For the fast-folding proteins, we use the fraction of native contacts RC \cite{best2013native}, which is a unitless measure of how many residues are within a cutoff distance to their folded-state neighbors.
	For Ala9, we use the $\rm{HB}_4$ coordinate, which measures the mean length of hydrogen bonds between residues $k$ and $k+4$ in the $\alpha$-helical structure \cite{ayaz_non-markovian_2021}.

	\begin{figure*}
		\vspace{-12mm}
		\includegraphics{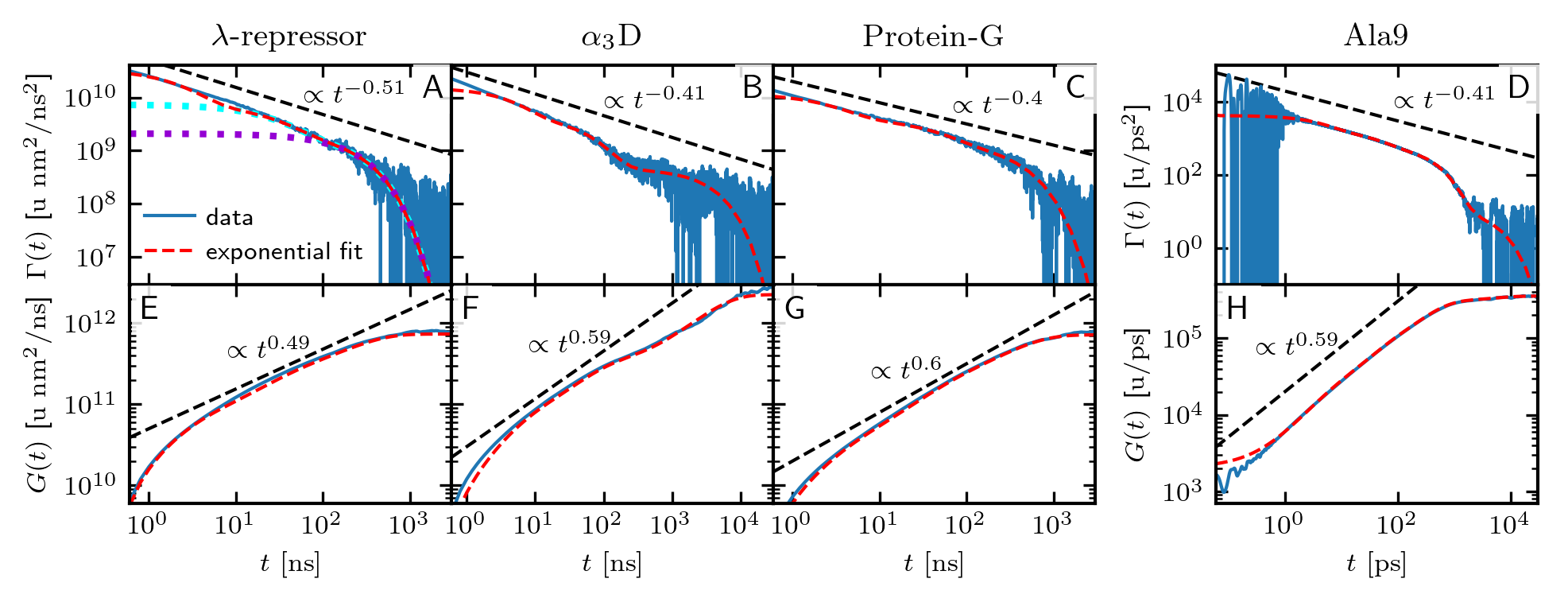}
		\caption{Friction kernels $\Gamma(t)$ (A)-(D) and their corresponding running integrals $G(t)$ (E)-(H), extracted from MD simulations for $\lambda$-repressor, $\alpha_3$D, protein-G \cite{lindorff-larsen_how_2011} and Ala9 \cite{ayaz_non-markovian_2021}, respectively (blue lines).
		The straight black dashed lines represent power-law behavior with values of $\alpha_{\rm{sub}}^{\rm{data}}$, extracted from the MSD as an average over Eq.~\eqref{eq_alpha} shown in Tab.~\ref{tab_alpha_protein}.
        The red dashed lines represent the multi-exponential fit to the friction kernel $\Gamma(t)$ in Eq.~\eqref{eq_kern_expo} using $n=5$ exponentials for Ala9 and $n=3$ exponentials for all other proteins.
		The dotted lines in (A) indicate contributions from the individual exponential components in $\Gamma(t)$ (Eq.~\eqref{eq_kern_expo}).
		The dashed red lines in (E)-(H) result from integration of the corresponding red dashed lines in (A)-(D), according to Eq.~\eqref{eq_int_kern}.
		}
		\label{fig2_kernels}
	\end{figure*}
	
	We extract the friction kernel $\Gamma(t)$ in the GLE Eq.~\eqref{eq_gle_wom} from trajectories using previously established methods \cite{ayaz_non-markovian_2021,daldrop2018butane,dalton_fast_2023}.
	In Figs.~\ref{fig2_kernels}A-D, we show that, for all proteins, the friction kernels show scaling behavior $\Gamma(t)\propto t^{-\alpha_{\rm{sub}}}$ with $0.40\leq\alpha_{\rm{sub}}\leq0.51$ on intermediate times and exhibit exponential decay for long times.
	In fact, the friction kernels are well described by a multi-exponential fit of the form
	\begin{equation}
		\label{eq_kern_expo}
		\Gamma(t) = \sum_{i=1}^{n}\frac{\gamma_i}{\tau_i} e^{-t/\tau_i} \,,
	\end{equation}
	with $n=3$ components for the fast-folding proteins and $n=5$ components for the homo-peptide Ala9.
	All friction coefficients $\gamma_i$ and memory times $\tau_i$ are given in the SI.
    Importantly, we observe clear power-law behavior in $\Gamma(t)$ over two decades in time, even with the inclusion of only three exponential terms.

    The finite length of this power-law behavior becomes especially apparent for the integrated friction $G(t)$, which is defined as
	\begin{equation}
		\label{eq_int_kern}
		G(t) = \int_{0}^{t} \Gamma(s) ds\,.
	\end{equation}
	For a pure power-law friction kernel $\Gamma(t)\propto t^{-\alpha_{\rm{sub}}}$ one obtains $G(t)\propto t^{1-\alpha_{\rm{sub}}}$.
	In Figs.~\ref{fig2_kernels}E-H it is clear that for all proteins, $G(t)$ (blue lines) does not exhibit pronounced power-law behavior over all time scales and is much better described by the multi-exponential Eq.~\eqref{eq_kern_expo} (red broken lines).
	Deviations between $G(t)$ and power-law behavior are most prominent on the shortest and longest resolved time scales.
	In the long time limit, the integrated kernel reaches a plateau, which corresponds to the total friction
	\begin{equation}
		\label{eq_total_friction}
		\gamma_{\rm{tot}} = G(\infty) = \sum_{i=1}^{n} \gamma_i\,.
	\end{equation}
	We use $\gamma_{\rm{tot}}$ to define the inertial time scale $\tau_m = m/\gamma_{\rm{tot}}$, which, for proteins, is much smaller than the diffusion time $\tau_D=\gamma_{\rm{tot}}L^2/k_{\rm{B}}T$.
	Here, $\tau_D$ is the time that the RC needs to diffusive over the distance $L$ for a flat free energy profile, where $L$ is the distance in the RC space between the unfolded state and the barrier top, as depicted in Fig.~\ref{fig1}B.
    In Tab.~\ref{tab_alpha_protein}, we present the ratios of $\tau_m/\tau_D$ for the different proteins.
    These values are very small, indicating that the RC dynamics are strongly overdamped for all proteins.

	\begin{table*}
		\centering
		\hspace{-6mm}
		\begin{tabular}{ |c|c|c|c|c|c|c|c|c| }
			\hline
			&  &  &  &  &  &  &  & \vspace*{-3mm}\\
			Protein & $U_0/k_{\rm{B}}T$ & $\tau_m/\tau_D$ & $c$ & $d$ & $\alpha_{\rm{sub}}^{\rm{pred}}$ & $\alpha_{\rm{sub}}^{\rm{data}}$ & $\gamma_3/\gamma_2$ & $\gamma_2/\gamma_1$ \\
			\hline
			$\lambda$-repressor & 2.1 & $2.6\times10^{-7}$ & 12 & 2.7  & $0.59\pm 0.04$ & $0.51 \pm 0.02$ & 3.7 & 2.5 \\ 
			\hline
			$\alpha_3$D & 3.2 & $3.9\times10^{-8}$ & 21 & 5.3 & $0.44\pm 0.02$ & $0.41\pm 0.06$ & 8.2 & 4.4 \\ 
			\hline 
			Protein-G & 5.0 & $2.1\times10^{-7}$ & 14 & 5.1 & $0.38\pm0.03$ & $0.40\pm 0.04$ & 5.1 & 5.2 \\ 
			\hline
			Ala9 & 1.3 & $1.4\times10^{-9}$ & 9 & 4.0 & $0.38\pm 0.04$ & $0.41\pm 0.06$ & 3.5 & 5.5\\
			\hline
		\end{tabular}
		\caption{Free energy barrier height $U_0$ relative to the global minimum, ratio of inertial time and diffusion time $\tau_m/\tau_D$, mean values for memory time ratio $c$ and friction coefficient ratio $d$, as defined in Eq.~\eqref{eq_cd}, and friction coefficients $\gamma_{i+1}/\gamma_i$ defined in Eq.~\eqref{eq_kern_expo}.
		The subdiffusive scaling prediction $\alpha_{\rm{sub}}^{\rm{pred}}$ from Eq.~\eqref{eq_inverse_cd} is compared to the extracted scaling exponent of the MD data $\alpha_{\rm{sub}}^{\rm{data}}$.
		The latter is obtained as a logarithmic time average of $\alpha(t)$ defined in Eq.~\eqref{eq_alpha}, i.e., an average using exponentially spaced time points, in the regime $\tau_1 < t <\tau_2$ for $\lambda$-repressor, $\alpha_3$D, and protein-G, and in the regime $\tau_2< t <\tau_4$ for Ala9.
		The complete sets of extracted memory kernel parameters are given in the SI.}
		\label{tab_alpha_protein}
	\end{table*}

    For friction coefficients $\gamma_i$ and memory times $\tau_i$ that are hierarchically ordered and exponentially spaced with ratios $c$ and $d$ according to
    \begin{align}
        \label{eq_cd}
        \begin{split}
        \tau_i &= \tau_1 c^{i-1} \\
        \gamma_i &= \gamma_1 d^{i-1}
        \end{split}
    \end{align}
    the friction kernel in Eq.~\eqref{eq_kern_expo} appears to satisfy a power law $\Gamma(t)\propto t^{-\alpha_{\rm{sub}}}$ over a finite range of time, where the range increases with an increasing number of exponential components \cite{noauthor_advances_2012,klimek2024_theo}.
	Interestingly, we find that the structure of the fitted memory components matches rather closely the hierarchical structure defined in Eq.~\eqref{eq_cd}, as indicated by the almost constant ratios of friction coefficients in Tab.~\ref{tab_alpha_protein} (the complete set of parameters is given in the SI).
    This leads to an even spacing between individual memory components on the log-log scale shown by the cyan, violet and red lines in Fig.~\ref{fig2_kernels}A.
	The exponent in the friction kernel $\Gamma(t)\propto t^{-\alpha_{\rm{sub}}}$ translates directly into a subdiffusive scaling regime in the MSD, with $C_{\rm{MSD}}(t)\propto t^{\alpha_{\rm{sub}}}$ for intermediate times \cite{burov_critical_2008, klimek2024_theo,noauthor_advances_2012}.
	The time-dependent scaling exponent is defined as the logarithmic derivative of the MSD as
	\begin{equation}
		\label{eq_alpha}
		\alpha(t) = \dfrac{{\rm{d}} \ln (C_{\rm{MSD}}(t))}{{\rm{d}} \ln (t)}\,,
	\end{equation}
	where the intermediate subdiffusive scaling for multi-exponential memory with $c>d$ and $d>1$ at times $\tau_1<t<\tau_n$ is derived in \cite{klimek2024_theo} as
	\begin{equation}
		\label{eq_inverse_cd}
		\alpha_{\rm{sub}}^{\rm{pred}} = \frac{\ln (c/d)}{\ln(c)} = \frac{\ln\left(\left(\frac{\gamma_1 \tau_i}{\tau_1\gamma_i}\right)^{\frac{1}{i-1}}\right)}{\ln\left(\left(\frac{\tau_i}{\tau_1}\right)^{\frac{1}{i-1}}\right)}\,.
	\end{equation}
	
\begin{figure*}
		\hspace{-15mm}
		\includegraphics{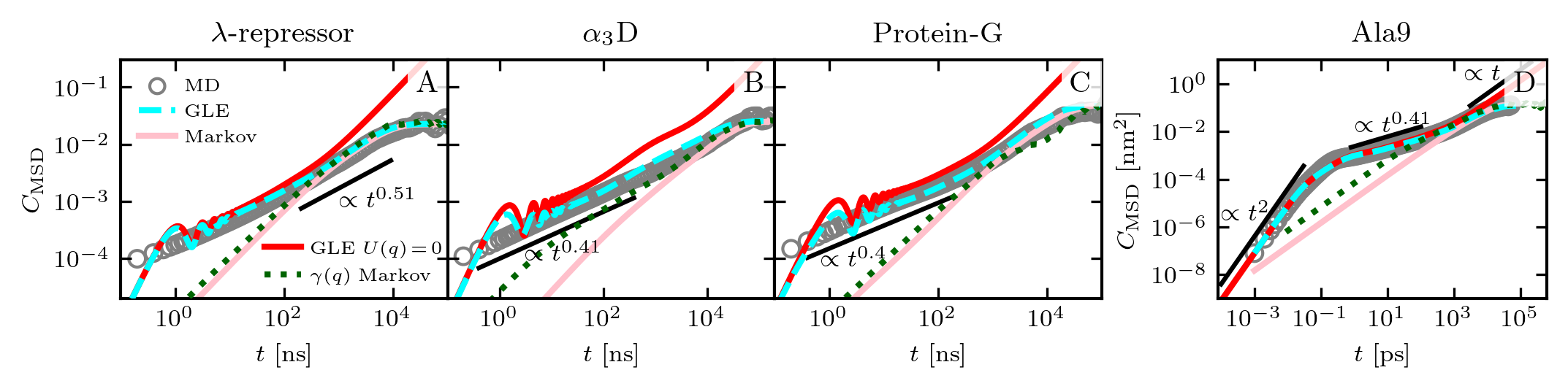}
		\caption{MSDs for different proteins.
		Data extracted from MD simulations \cite{lindorff-larsen_how_2011, ayaz_non-markovian_2021} are shown as gray circles.
        The cyan dashed lines represent the GLE simulation result using the fitted kernel $\Gamma(t)$ of Figs.~\ref{fig2_kernels}A-D and the free energy landscape $U(q)$ in Figs.~\ref{fig1}B-D.
        The red line is the analytical GLE prediction for $U(q)=0$ \cite{klimek2024_theo}.
        The dotted line results from simulation of Eq.~\eqref{eq_pos_dep_motion} with coordinate-dependent friction profiles $\gamma_{\rm{unf}}(q)$ extracted from the unfolding MFPT profile of the MD data (shown in the SI).
        The pink lines originate from simulations of the Markovian limit using the constant total friction $\gamma_{\rm{tot}}$ and $U(q)$.
        Simulation details are given in the SI.}
		\label{fig3_protein_msd}
	\end{figure*}

    In Tab.~\ref{tab_alpha_protein}, we compare the subdiffusive scaling exponent $\alpha_{\rm{sub}}^{\rm{pred}}$ to the observed subdiffusive scaling in the MSD $\alpha_{\rm{sub}}^{\rm{data}}$.
    The former is predicted by Eq.~\eqref{eq_inverse_cd} using friction components $\gamma_i$ and memory times $\tau_i$ with $i\leq3$ from the fit of the memory kernels in Fig.~\ref{fig2_kernels}; the latter is computed as a logarithmic time average of $\alpha(t)$ in the subdiffusive regime using Eq.~\eqref{eq_alpha}.
	The observed scaling exponents $\alpha_{\rm{sub}}^{\rm{data}}$ agree well with the scaling exponent $\alpha_{\rm{sub}}^{\rm{pred}}$, where
	the prediction of the subdiffusive scaling via Eq.~\eqref{eq_inverse_cd} does not include effects due to the free energy landscape \cite{klimek2024_theo}.
	This suggests that the memory alone dictates the subdiffusive scaling in the MSD of fast-folding protein dynamics.
	In Fig.~\ref{fig3_protein_msd}, we show the MSD of all proteins from MD simulations as gray circles.
	The MSD from simulations of the GLE Eq.~\eqref{eq_gle_wom} (cyan broken line) describes the MSD from MD data very well for all proteins on a double-logarithmic scale.
    Here, it should be noted that the oscillations on short times in the MSD from GLE simulations are due to unresolved memory effects at times below the temporal resolution of the available data, as we show later.
	For simulations of the GLE, we use the extracted free energy landscape $U(q)$ Fig.~\ref{fig1}B-E, the fitted multi-exponential friction kernels Fig.~\ref{fig2_kernels}A-D and a Markovian embedding scheme \cite{simulation_git2025,noauthor_advances_2012,ayaz_non-markovian_2021,brunig_time-dependent_2022}.
	For Ala9 in Fig.~\ref{fig3_protein_msd}D, the temporal resolution is higher ($\Delta=1\,\rm{fs}$ compared to $\Delta=0.2\,\rm{ns}$ for the other proteins), which leads to a perfect agreement between the GLE simulation and the MD data for all times, even resolving the short-time ballistic regime, where $\alpha=2$.
	
	Interestingly, simulations in the Markovian limit, i.e. using ${\Gamma(t)=2\gamma_{\rm{tot}}\delta(t)}$ and the free energy profiles $U(q)$ extracted from the MD simulations, do not reproduce the subdiffusive scaling behavior at intermediate times (pink lines in Fig.~\ref{fig3_protein_msd}).
    Overall, the MSD for these Markovian simulations deviates strongly from the MSD from the MD data for all proteins.
	Even when generalizing the overdamped Langevin equation to include coordinate-dependent friction $\gamma(q)$ according to \cite{hinczewski_how_2010}
	\begin{equation}
		\label{eq_pos_dep_motion}
		0 = -\nabla U(q) - \gamma(q) \dot{q} - \frac{k_{\rm{B}}T}{2}\frac{\gamma'(q)}{\gamma(q)}+ \sqrt{k_{\rm{B}}T \gamma(q)} \xi(t)\,,
	\end{equation}
	the resulting MSDs (dotted lines in Fig.~\ref{fig3_protein_msd}) do not exhibit the correct scaling behavior, deviating strongly from the MD MSDs.
	In the SI we show that the inclusion of inertial effects in Eq.~\eqref{eq_pos_dep_motion} also does not improve the agreement with the MD results.
	
\begin{figure}
		\vspace{-0mm}
		\hspace{-2mm}
		\includegraphics{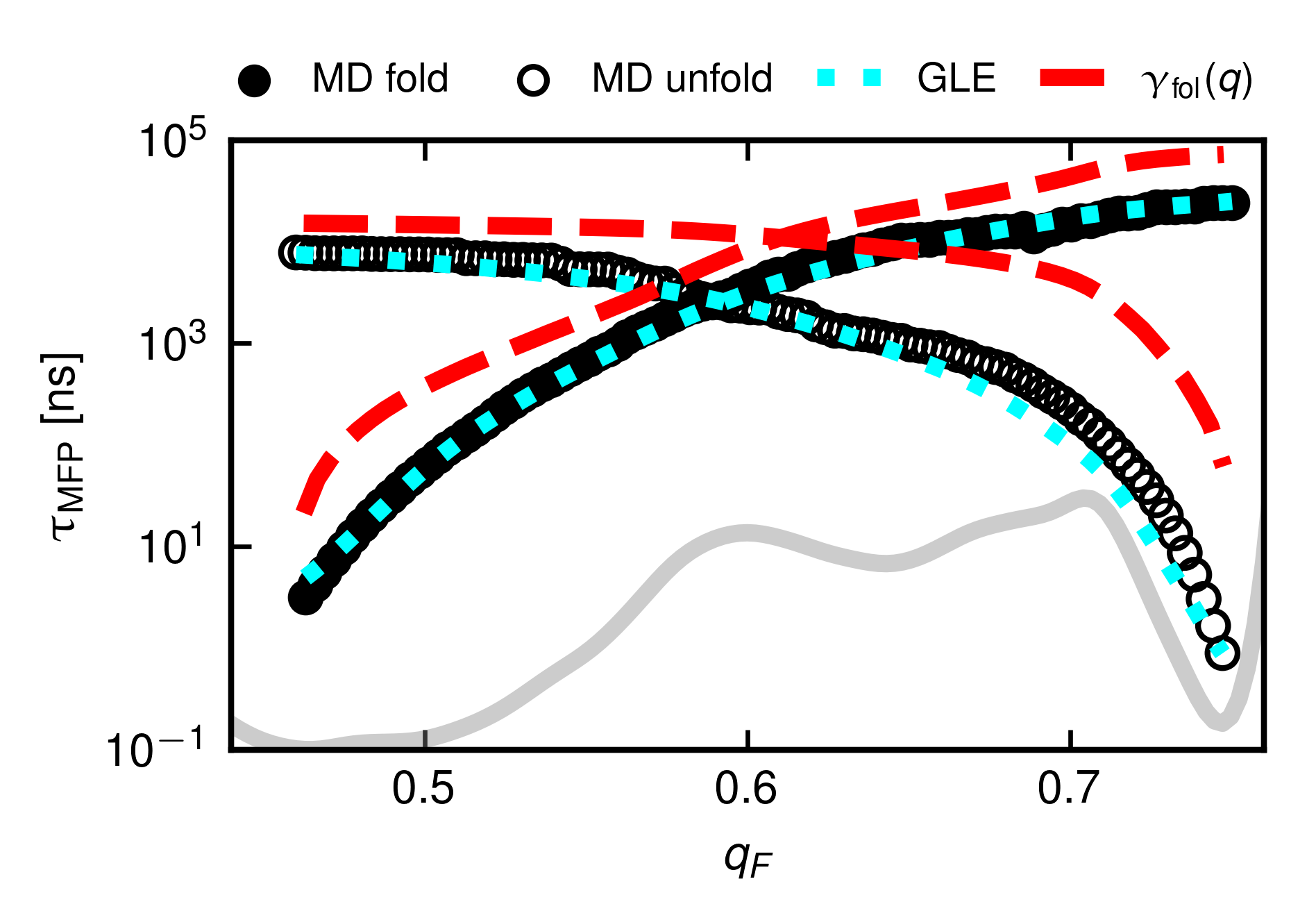}
		\caption{
			MFPTs from MD simulations of $\lambda$-repressor in the folding direction, starting from $q_S=0.38$ (black dots) and in the unfolding direction starting from $q_S=0.75$ (black circles).
			The gray line shows the shape of the free energy profile $U(q)$.
            The predictions in the respective directions are shown in cyan for the GLE with explicit memory effects, and in red for the coordinate-dependent friction profile $\gamma_{\rm{fol}}(q)$ extracted from the MFPT in the folding direction.
			See SI for details on MFPTs and $\gamma(q)$ extraction.
		}
		\label{fig_lambda_mfpt}
	\end{figure}

	The GLE captures not only the MSD but also the MFPT profiles with high accuracy, as exemplified in Fig.~\ref{fig_lambda_mfpt} for $\lambda$-repressor.
	In contrast, the folding dynamics are not consistently captured by a Markovian model with coordinate-dependent friction defined in Eq.~\eqref{eq_pos_dep_motion}, as shown in Fig.~\ref{fig_lambda_mfpt}: the dashed red lines deviate significantly from the MFPT of the MD data (black circles).
    Additional details on all MFPT profiles and coordinate-dependent friction profiles $\gamma(q)$ are given in the SI for all proteins.

\begin{figure}
		\vspace{-0mm}
		\hspace{-2mm}
		\includegraphics{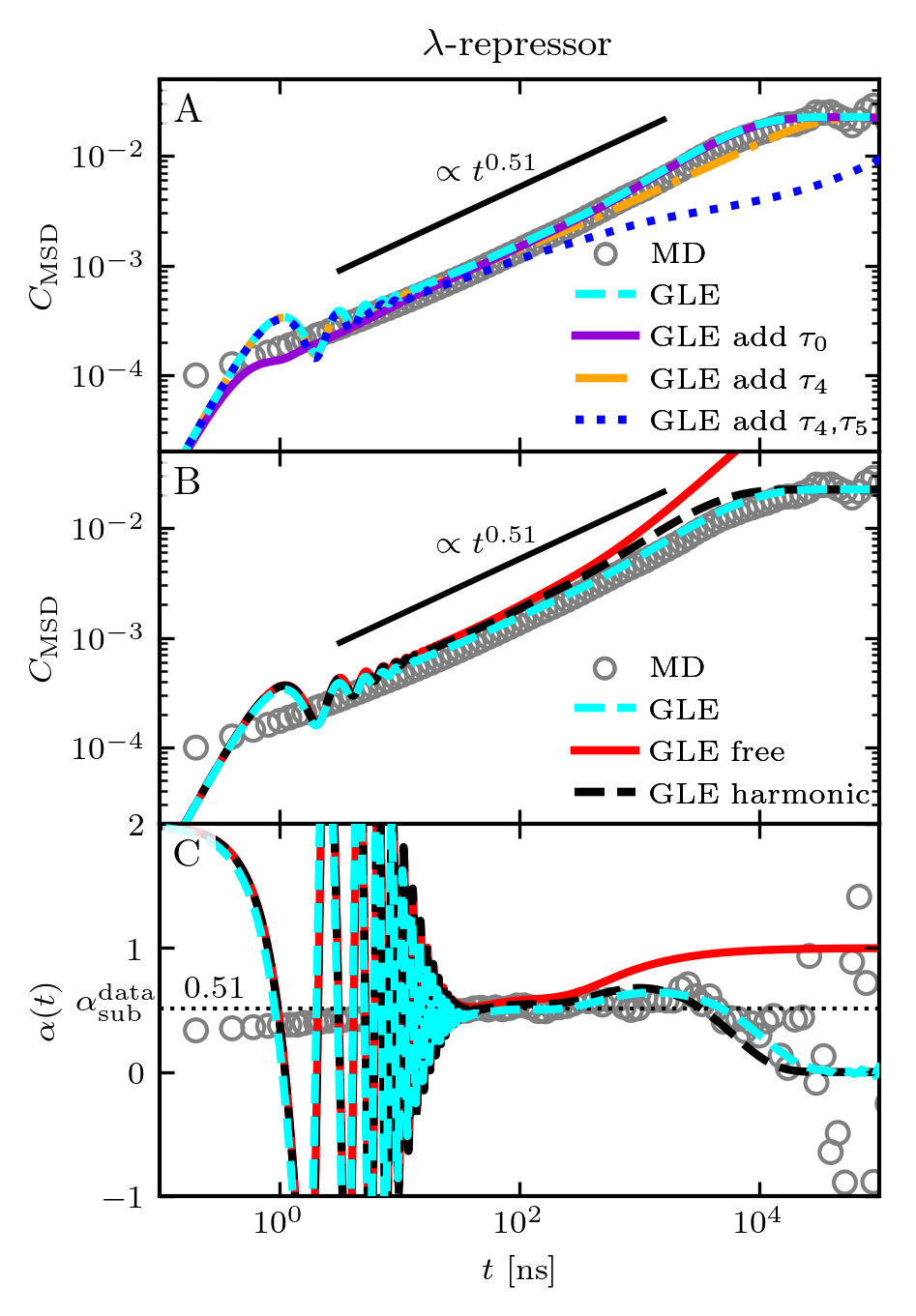}
		\caption{$\lambda$-repressor MSD and scaling exponent modeled by the GLE.
		(A) MSD from MD simulations \cite{lindorff-larsen_how_2011} (gray circles) compared to various GLE simulations.
		The dashed cyan line represents the result with the extracted $U(q)$ and fitted $\Gamma(t)$.
		The violet, orange and blue lines result from GLE simulations with additional memory components on shorter time scale $\tau_0$, longer time scale $\tau_4$ and two longer time scales $\tau_4$, $\tau_5$, respectively, according to Eq.~\eqref{eq_kern_expo} with values of $c$ and $d$ in Eq.~\eqref{eq_cd} taken from Tab.~\ref{tab_alpha_protein}.
		(B) The black dashed line is the analytical result for the MSD in a harmonic potential $U(q)=Kq^2/2$ with $K=k_BT/\langle (q -\langle q \rangle)^2 \rangle$.
		(C) Time-dependent exponent extracted by Eq.~\eqref{eq_alpha} from the lines shown in (B).
        The horizontal dotted line shows the average of the MD data, which is $\alpha_{\rm{sub}}^{\rm{data}}=0.51$ given in Tab.~\ref{tab_alpha_protein}.
		}
		\label{fig4_lambda_msd}
	\end{figure}

	For short times, the MSDs from GLE simulations, parameterized by the fitted kernels, exhibit nanosecond-scale oscillations (Figs.~\ref{fig3_protein_msd}A-C), which are absent in the MD MSD.
	These oscillations are reduced by adding an exponential component to the memory kernel (Eq.~\eqref{eq_kern_expo}) at shorter times, where $\tau_i$ and $\gamma_i$ of the additional term are extrapolated by the average ratios $c$ and $d$, defined in Eq.~\eqref{eq_cd} and given in Tab.~\ref{tab_alpha_protein}.
	This reduction of oscillations is seen by comparing the violet and cyan lines in Fig.~\ref{fig4_lambda_msd}A and it further improves the agreement with the MSD of the MD data.
	The addition of memory components on shorter time scales leaves the total friction $\gamma_{\rm{tot}}$ virtually unchanged, since $c>1$ and $d>1$ lead to exponentially smaller friction coefficients for the short memory times.
    The improvement of the MSD description while conserving the total friction suggests that there are likely more exponential components present on time scales below the temporal resolution of the data (see SI for results for all other proteins).
	This further indicates that the hierarchy of the memory components, Eq.~\eqref{eq_cd}, is appropriate to describe protein conformational dynamics.
	On the contrary, adding memory components according to the observed hierarchy on longer time scales drastically increases $\gamma_{\rm{tot}}$ and leads to deviations compared to the MSD of the MD data.
    While adding one memory component on a longer time scale (orange line in Fig.~\ref{fig4_lambda_msd}A) leads to a comparable deviation from the MD data as the GLE prediction (cyan line), the addition of two memory components on longer time scales (dark-blue line) clearly leads to strong deviations from the MD data (more detail in the SI).
	This implies that the longest time scale of the memory is resolved by our fit of $\Gamma(t)$, whereas shorter memory times are involved in the fast-folding protein dynamics that are not resolved due to the finite time resolution of the available data.
    
    Deviations between the analytically predicted MSD from the GLE for $U(q)=0$ \cite{klimek2024_theo} and the MSD of the MD data increase for increasing barrier height $U_0$ (see Tab.~\ref{tab_alpha_protein}), as shown in Fig.~\ref{fig3_protein_msd}.
	For $\lambda$-repressor and Ala9, with maximal barrier heights of $2.1k_{\rm{B}}T$ and $1.3k_{\rm{B}}T$, respectively, the dynamics are dictated by memory alone and the main influence of the free energy landscape is the long-time confinement.
	Therefore, the analytical result of the MSD \cite{klimek2024_theo} with multi-exponential memory Eq.~\eqref{eq_kern_expo} in a harmonic potential that leads to the same positional variance as the free energy $U(q)$, captures the dynamics of $\lambda$-repressor and Ala9 well, as shown for $\lambda$-repressor by the black dashed line in Fig.~\ref{fig4_lambda_msd}B.
	Moreover, the time dependent scaling behavior $\alpha(t)$ of the MD data extracted via Eq.~\eqref{eq_alpha} (gray circles) is well captured by the description of the GLE with a harmonic potential (dashed black line), as shown in Fig.~\ref{fig4_lambda_msd}C for $\lambda$-repressor.
	For higher free energy barriers, as exhibited by $\alpha_3$D ($3.2\,k_{\rm{B}}T$) and protein-G ($5.0\,k_{\rm{B}}T$), the MSD for the analytical prediction with $U(q)=0$ appears shifted upwards compared to the MD result, i.e. the red lines are higher than the gray circles in Fig.~\ref{fig3_protein_msd}B,C at intermediate times.
    Thus, the accurate prediction of the MSD of $\alpha_3$D and protein-G requires both, the knowledge of the free energy and the memory.
    Nevertheless, the intermediate subdiffusive scaling is still captured well by Eq.~\eqref{eq_inverse_cd} (shown in Tab.~\ref{tab_alpha_protein}), which includes no information about $U(q)$.
    This further implies that memory is the main influence on the subdiffusion in the protein folding dynamics compared to energy-barrier effects.
	
	\section{Discussion \& Conclusions}
	We extract the friction memory kernel $\Gamma(t)$ of different fast-folding proteins from MD simulations and find that each protein is well described by multi-exponential $\Gamma(t)$ as shown in Fig.~\ref{fig2_kernels}.
	In previous work, it has been suggested that protein folding is fractal and thus exhibits power-law behavior and subdiffusion across all scales \cite{min2005observation,hu_dynamics_2016,kou_generalized_2004,sangha_proteins_2009}.
    However, it has also been argued that proteins fold hierarchically \cite{baldwin_is_1999,mori_molecular_2016}.
	We show that fast-folding protein conformational dynamics are well described by the sum of a small number ($n=3-5$) of exponential memory components, which leads to effective power-law behavior in the memory kernel and the MSD over a finite time range.
	In fact, the memory components follow a hierarchical pattern with approximately constant ratios of friction coefficients and memory times (see Tab.~\ref{tab_alpha_protein} and the SI) Eq.~\eqref{eq_cd}, which gives rise to a subdiffusive regime in the MSD.
    The hierarchical structure of the memory in the fraction of native contacts RC is possibly related to hierarchical folding mechanisms.
    This would mean that individual memory components are related to the dynamics of differently sized subunits of the protein which move on different time scales.
    In order to test this, one could use amino acid mutation studies, coarse grained simulation methods or MD simulations for different conditions to connect the memory contributions to structural and environmental features.
    For instance, it was recently shown that by reducing pH and thereby eliminating salt-bridge interactions between tertiary structures, contributions from the long timescale friction components are dramatically reduced \cite{dalton2024ph}.
	Moreover, the description of subdiffusive behavior by multi-exponential friction memory in the GLE is transferable to other systems that exhibit subdiffusion, for instance to cell motion \cite{dieterich_anomalous_2008,klimek_2024_cell,klimek_data-driven_2024} or to polymer network dynamics \cite{guo_entanglement-controlled_2012,abbasi_non-markovian_2023}.
	
	The scaling exponent $\alpha_{\rm{sub}}^{\rm{data}}$ in the subdiffusive MSD regime is accurately predicted by $c$ and $d$ according to Eq.~\eqref{eq_inverse_cd}, which is independent of the free energy landscape.
	Moreover, the MSD from MD simulations is accurately predicted by the description with the GLE, as shown in Fig.~\ref{fig3_protein_msd} and the behavior of MFPT profiles is captured by the GLE as well, as shown in Fig.~\ref{fig_lambda_mfpt} and in the SI.
	We can estimate time scales of memory components below the temporal resolution of the data by extrapolating the hierarchical pattern of the memory components to shorter time scales.
    The inclusion of memory components on these shorter time scales yields a more accurate description of the MSD, shown in Fig.~\ref{fig4_lambda_msd}.
	In contrast, the addition of memory contributions on longer time scales worsens the agreement with the MSD of the MD data.
	This suggests that the upper temporal bound for dynamical processes of the proteins, and consequently the longest memory time scale, are correctly captured by the data, whereas there are more time scales involved in the dynamics that are below the temporal resolution of the data.
    Previously developed optimization methods \cite{tepper2024accurate} represent an alternative method to determine time scales below the temporal resolution of the data that is independent of the observed ratios $c$ and $d$.

    Memory effects dominate the subdiffusive behavior of fast-folding proteins, more so than contributions from the free energy landscape.
	This is seen by comparing GLE simulations with and without memory in the free energy landscape $U(q)$ (Fig.~\ref{fig3_protein_msd}).
    Without memory, the MSD does not exhibit subdiffusion.
	For proteins with small free energy barriers up to $\sim 2k_{\rm{B}}T$, the MSD is well captured by the description with the GLE and $U(q)=0$, except for the long time confinement that is present in spatially bounded free energy landscapes (Fig.~\ref{fig3_protein_msd}).
    This underscores the importance of memory effects on the dynamics compared to effects due to the free energy landscape.
	We expect that for proteins with significantly higher free energy barriers than those considered in this study, memory effects are still crucial to describe the dynamics but additional subdiffusive regimes in the MSD can arise from barrier crossing processes \cite{klimek2024_theo,goychuk_finite-range_2020,evers_colloids_2013}.
	For short times, when the dynamics have not yet coupled to the local shape of the free energy landscape, memory dominates protein conformational dynamics, as the dynamics are usually highly overdamped, i.e. $\tau_m \ll \tau_D$ as shown in Tab.~\ref{tab_alpha_protein} and explained in \cite{klimek2024_theo}.
	
	Neither the Markovian limit nor the assumption of coordinate-dependent Markovian friction $\gamma(q)$ Eq.~\eqref{eq_pos_dep_motion} reproduce subdiffusive scaling in the MSD (Figs.~\ref{fig3_protein_msd}, \ref{fig4_lambda_msd}).
	The coordinate-dependent friction $\gamma(q)$ can also not consistently describe MFPT profiles (Fig.~\ref{fig_lambda_mfpt} and SI).
	Overall, the GLE accurately describes the RC dynamics of protein folding, whereas Markovian models fail to capture the dynamics.
	
	Our findings highlight the importance of memory effects in protein conformational dynamics.
	In line with previous studies \cite{dalton_fast_2023}, we find that friction memory effects dominate the subdiffusive RC dynamics of fast-folding proteins compared to the effects of the free energy landscape.
	Our results imply that fast-folding proteins are well described by exponential memory contributions with exponentially spaced memory times and friction coefficients that possibly reflect hierarchical folding dynamics.
	The hierarchically structured memory accurately captures the finite-time fractal behavior in fast-folding protein dynamics for intermediate times and showcases the presence of a longest time scale in the dynamics.

    \section*{Acknowledgements}
    We acknowledge funding by the Deutsche Forschungsgemeinschaft (DFG) through grant CRC 1449 “Dynamic Hydrogels at Biointerfaces”, Project ID 431232613, Project A03.

    \section*{Data availability}
    The code to simulate the GLE and position dependent LE are available at \url{https://github.com/kanton42/msd_subdiffusion} and kernel extraction methods at \url{https://github.com/lucastepper/memtools}.

	\bibliographystyle{unsrt}
	\bibliography{msd}

\begin{thebibliography}{10}

\bibitem{jumper2021highly}
John Jumper, Richard Evans, Alexander Pritzel, Tim Green, Michael Figurnov,
  Olaf Ronneberger, Kathryn Tunyasuvunakool, Russ Bates, Augustin
  {\v{Z}}{\'\i}dek, Anna Potapenko, et~al.
\newblock Highly accurate protein structure prediction with alphafold.
\newblock {\em nature}, 596(7873):583--589, 2021.

\bibitem{sweeney2017protein}
Patrick Sweeney, Hyunsun Park, Marc Baumann, John Dunlop, Judith Frydman, Ron
  Kopito, Alexander McCampbell, Gabrielle Leblanc, Anjli Venkateswaran, Antti
  Nurmi, et~al.
\newblock Protein misfolding in neurodegenerative diseases: implications and
  strategies.
\newblock {\em Translational neurodegeneration}, 6:1--13, 2017.

\bibitem{fukui_formulation_1970}
Kenichi Fukui.
\newblock Formulation of the reaction coordinate.
\newblock {\em J. Phys. Chem.}, 74(23):4161--4163, November 1970.

\bibitem{best_reaction_2005}
Robert~B. Best and Gerhard Hummer.
\newblock Reaction coordinates and rates from transition paths.
\newblock {\em Proceedings of the National Academy of Sciences},
  102(19):6732--6737, May 2005.
\newblock Publisher: Proceedings of the National Academy of Sciences.

\bibitem{kmiecik2016coarse}
Sebastian Kmiecik, Dominik Gront, Michal Kolinski, Lukasz Wieteska,
  Aleksandra~Elzbieta Dawid, and Andrzej Kolinski.
\newblock Coarse-grained protein models and their applications.
\newblock {\em Chemical reviews}, 116(14):7898--7936, 2016.

\bibitem{heath2007coarse}
Allison~P Heath, Lydia~E Kavraki, and Cecilia Clementi.
\newblock From coarse-grain to all-atom: toward multiscale analysis of protein
  landscapes.
\newblock {\em Proteins: Structure, Function, and Bioinformatics},
  68(3):646--661, 2007.

\bibitem{monago2025unraveling}
Carlos Monago, JA~de~la Torre, R~Delgado-Buscalioni, and Pep Espa{\~n}ol.
\newblock Unraveling internal friction in a coarse-grained protein model.
\newblock {\em The Journal of Chemical Physics}, 162(11), 2025.

\bibitem{krivov_reaction_2013}
Sergei~V. Krivov.
\newblock On {Reaction} {Coordinate} {Optimality}.
\newblock {\em J. Chem. Theory Comput.}, 9(1):135--146, January 2013.
\newblock Publisher: American Chemical Society.

\bibitem{kramers_brownian_1940}
H.~A. Kramers.
\newblock Brownian motion in a field of force and the diffusion model of
  chemical reactions.
\newblock {\em Physica}, 7(4):284--304, April 1940.

\bibitem{shukla2015markov}
Diwakar Shukla, Carlos~X Hern{\'a}ndez, Jeffrey~K Weber, and Vijay~S Pande.
\newblock Markov state models provide insights into dynamic modulation of
  protein function.
\newblock {\em Accounts of chemical research}, 48(2):414--422, 2015.

\bibitem{malmstrom2014application}
Robert~D Malmstrom, Christopher~T Lee, Adam~T Van~Wart, and Rommie~E Amaro.
\newblock Application of molecular-dynamics based markov state models to
  functional proteins.
\newblock {\em Journal of chemical theory and computation}, 10(7):2648--2657,
  2014.

\bibitem{ayaz_non-markovian_2021}
Cihan Ayaz, Lucas Tepper, Florian~N. Brünig, Julian Kappler, Jan~O. Daldrop,
  and Roland~R. Netz.
\newblock Non-{Markovian} modeling of protein folding.
\newblock {\em Proceedings of the National Academy of Sciences},
  118(31):e2023856118, August 2021.
\newblock Publisher: Proceedings of the National Academy of Sciences.

\bibitem{dalton_fast_2023}
Benjamin~A. Dalton, Cihan Ayaz, Henrik Kiefer, Anton Klimek, Lucas Tepper, and
  Roland~R. Netz.
\newblock Fast protein folding is governed by memory-dependent friction.
\newblock {\em Proceedings of the National Academy of Sciences},
  120(31):e2220068120, August 2023.
\newblock Publisher: Proceedings of the National Academy of Sciences.

\bibitem{min2005observation}
Wei Min, Guobin Luo, Binny~J Cherayil, SC~Kou, and X~Sunney Xie.
\newblock Observation of a power-law memory kernel for fluctuations within a
  single protein molecule.
\newblock {\em Physical review letters}, 94(19):198302, 2005.

\bibitem{kou_generalized_2004}
S.~C. Kou and X.~Sunney Xie.
\newblock Generalized {Langevin} {Equation} with {Fractional} {Gaussian}
  {Noise}: {Subdiffusion} within a {Single} {Protein} {Molecule}.
\newblock {\em Phys. Rev. Lett.}, 93(18):180603, October 2004.
\newblock Publisher: American Physical Society.

\bibitem{plotkin_non-markovian_1998}
Steven~S. Plotkin and Peter~G. Wolynes.
\newblock Non-{Markovian} {Configurational} {Diffusion} and {Reaction}
  {Coordinates} for {Protein} {Folding}.
\newblock {\em Phys. Rev. Lett.}, 80(22):5015--5018, June 1998.
\newblock Publisher: American Physical Society.

\bibitem{berezhkovskii_single-molecule_2018}
Alexander~M. Berezhkovskii and Dmitrii~E. Makarov.
\newblock Single-{Molecule} {Test} for {Markovianity} of the {Dynamics} along a
  {Reaction} {Coordinate}.
\newblock {\em J. Phys. Chem. Lett.}, 9(9):2190--2195, May 2018.
\newblock Publisher: American Chemical Society.

\bibitem{vollmar_model-free_2024}
Leonie Vollmar, Rick Bebon, Julia Schimpf, Bastian Flietel, Sirin Celiksoy,
  Carsten Sönnichsen, Aljaž Godec, and Thorsten Hugel.
\newblock Model-free inference of memory in conformational dynamics of a
  multi-domain protein, April 2024.
\newblock arXiv:2404.16799 [cond-mat, physics:physics].

\bibitem{dalton2024ph}
Benjamin~A Dalton and Roland~R Netz.
\newblock ph modulates friction memory effects in protein folding.
\newblock {\em Physical Review Letters}, 133(18):188401, 2024.

\bibitem{mori_transport_1965}
Hazime Mori.
\newblock Transport, {Collective} {Motion}, and {Brownian} {Motion}.
\newblock {\em Prog. Theor. Phys.}, 33(3):423--455, March 1965.

\bibitem{ayaz_generalized_2022}
Cihan Ayaz, Laura Scalfi, Benjamin~A. Dalton, and Roland~R. Netz.
\newblock Generalized {Langevin} equation with a nonlinear potential of mean
  force and nonlinear memory friction from a hybrid projection scheme.
\newblock {\em Phys. Rev. E}, 105(5):054138, May 2022.
\newblock Publisher: American Physical Society.

\bibitem{dalton2024memory}
Benjamin~A Dalton, Anton Klimek, Henrik Kiefer, Florian~N Br{\"u}nig,
  H{\'e}l{\`e}ne Colinet, Lucas Tepper, Amir Abbasi, and Roland~R Netz.
\newblock Memory and friction: From the nanoscale to the macroscale.
\newblock {\em Annual Review of Physical Chemistry}, 76, 2024.

\bibitem{vroylandt2022derivation}
Hadrien Vroylandt.
\newblock On the derivation of the generalized langevin equation and the
  fluctuation-dissipation theorem.
\newblock {\em Europhysics Letters}, 140(6):62003, 2022.

\bibitem{best2010coordinate}
Robert~B Best and Gerhard Hummer.
\newblock Coordinate-dependent diffusion in protein folding.
\newblock {\em Proceedings of the National Academy of Sciences},
  107(3):1088--1093, 2010.

\bibitem{hinczewski_how_2010}
M.~Hinczewski, Y.~von Hansen, J.~Dzubiella, and R.~R. Netz.
\newblock How the diffusivity profile reduces the arbitrariness of protein
  folding free energies.
\newblock {\em The Journal of Chemical Physics}, 132(24):245103, June 2010.

\bibitem{metzler_random_2000}
Ralf Metzler and Joseph Klafter.
\newblock The random walk's guide to anomalous diffusion: a fractional dynamics
  approach.
\newblock {\em Physics Reports}, 339(1):1--77, December 2000.

\bibitem{kappler_cyclization_2019}
Julian Kappler, Frank Noé, and Roland~R. Netz.
\newblock Cyclization and {Relaxation} {Dynamics} of {Finite}-{Length}
  {Collapsed} {Self}-{Avoiding} {Polymers}.
\newblock {\em Phys. Rev. Lett.}, 122(6):067801, February 2019.
\newblock Publisher: American Physical Society.

\bibitem{noauthor_advances_2012}
Igor Goychuk.
\newblock {\em Advances in {Chemical} {Physics}}.
\newblock John Wiley \& Sons, Ltd, 1 edition, 2012.

\bibitem{klimek2024_theo}
Anton Klimek, Benjamin~A Dalton, and Roland~R Netz.
\newblock Subdiffusion from competition between multi-exponential friction
  memory and energy barriers.
\newblock {\em arXiv preprint arXiv:2506.03036}, 2025.

\bibitem{hartl2017protein}
F~Ulrich Hartl.
\newblock Protein misfolding diseases.
\newblock {\em Annual review of biochemistry}, 86(1):21--26, 2017.

\bibitem{lindorff-larsen_how_2011}
Kresten Lindorff-Larsen, Stefano Piana, Ron~O. Dror, and David~E. Shaw.
\newblock How {Fast}-{Folding} {Proteins} {Fold}.
\newblock {\em Science}, 334(6055):517--520, October 2011.
\newblock Publisher: American Association for the Advancement of Science.

\bibitem{baldwin_is_1999}
Robert~L. Baldwin and George~D. Rose.
\newblock Is protein folding hierarchic? {I}. {Local} structure and peptide
  folding.
\newblock {\em Trends in Biochemical Sciences}, 24(1):26--33, January 1999.

\bibitem{mori_molecular_2016}
Toshifumi Mori and Shinji Saito.
\newblock Molecular {Mechanism} {Behind} the {Fast} {Folding}/{Unfolding}
  {Transitions} of {Villin} {Headpiece} {Subdomain}: {Hierarchy} and
  {Heterogeneity}.
\newblock {\em J. Phys. Chem. B}, 120(45):11683--11691, November 2016.
\newblock Publisher: American Chemical Society.

\bibitem{sangha_proteins_2009}
Amandeep~K. Sangha and T.~Keyes.
\newblock Proteins {Fold} by {Subdiffusion} of the {Order} {Parameter}.
\newblock {\em J. Phys. Chem. B}, 113(48):15886--15894, December 2009.
\newblock Publisher: American Chemical Society.

\bibitem{satija_transition_2017}
Rohit Satija, Atanu Das, and Dmitrii~E. Makarov.
\newblock Transition path times reveal memory effects and anomalous diffusion
  in the dynamics of protein folding.
\newblock {\em J. Chem. Phys.}, 147(15):152707, October 2017.

\bibitem{hu_dynamics_2016}
Xiaohu Hu, Liang Hong, Micholas Dean~Smith, Thomas Neusius, Xiaolin Cheng, and
  Jeremy~C. Smith.
\newblock The dynamics of single protein molecules is non-equilibrium and
  self-similar over thirteen decades in time.
\newblock {\em Nature Phys}, 12(2):171--174, February 2016.
\newblock Number: 2 Publisher: Nature Publishing Group.

\bibitem{englander2014nature}
S~Walter Englander and Leland Mayne.
\newblock The nature of protein folding pathways.
\newblock {\em Proceedings of the National Academy of Sciences},
  111(45):15873--15880, 2014.

\bibitem{thirumalai2001chaperonin}
Devarajan Thirumalai and George~H Lorimer.
\newblock Chaperonin-mediated protein folding.
\newblock {\em Annual review of biophysics and biomolecular structure},
  30(1):245--269, 2001.

\bibitem{stayrook2008crystal}
Steven Stayrook, Peera Jaru-Ampornpan, Jenny Ni, Ann Hochschild, and Mitchell
  Lewis.
\newblock Crystal structure of the $\lambda$ repressor and a model for pairwise
  cooperative operator binding.
\newblock {\em Nature}, 452(7190):1022--1025, 2008.

\bibitem{bjorck1984purification}
L~Bj{\"o}rck and G~Kronvall.
\newblock Purification and some properties of streptococcal protein g, a novel
  igg-binding reagent.
\newblock {\em Journal of Immunology (Baltimore, Md.: 1950)}, 133(2):969--974,
  1984.

\bibitem{zhu2003ultrafast}
Yongjin Zhu, Darwin~OV Alonso, Kosuke Maki, Cheng-Yen Huang, Steven~J Lahr,
  Valerie Daggett, Heinrich Roder, William~F DeGrado, and Feng Gai.
\newblock Ultrafast folding of $\alpha$3d: A de novo designed three-helix
  bundle protein.
\newblock {\em Proceedings of the National Academy of Sciences},
  100(26):15486--15491, 2003.

\bibitem{kuczera2024helix}
Krzysztof Kuczera, Gouri~S Jas, and Robert Szoszkiewicz.
\newblock Helix formation from hydrogen bond kinetics in alanine homopeptides.
\newblock {\em Crystals}, 14(6):532, 2024.

\bibitem{tepper2024accurate}
Lucas Tepper, Benjamin Dalton, and Roland~R Netz.
\newblock Accurate memory kernel extraction from discretized time-series data.
\newblock {\em Journal of Chemical Theory and Computation}, 20(8):3061--3068,
  2024.

\bibitem{best2013native}
Robert~B Best, Gerhard Hummer, and William~A Eaton.
\newblock Native contacts determine protein folding mechanisms in atomistic
  simulations.
\newblock {\em Proceedings of the National Academy of Sciences},
  110(44):17874--17879, 2013.

\bibitem{daldrop2018butane}
Jan~O Daldrop, Julian Kappler, Florian~N Br{\"u}nig, and Roland~R Netz.
\newblock Butane dihedral angle dynamics in water is dominated by internal
  friction.
\newblock {\em Proceedings of the National Academy of Sciences},
  115(20):5169--5174, 2018.

\bibitem{burov_critical_2008}
S.~Burov and E.~Barkai.
\newblock Critical {Exponent} of the {Fractional} {Langevin} {Equation}.
\newblock {\em Phys. Rev. Lett.}, 100(7):070601, February 2008.
\newblock Publisher: American Physical Society.

\bibitem{simulation_git2025}
Anton Klimek.
\newblock Simulation tools and implementations for multi-exponential memory,
  2025.

\bibitem{brunig_time-dependent_2022}
Florian~N. Brünig, Otto Geburtig, Alexander~von Canal, Julian Kappler, and
  Roland~R. Netz.
\newblock Time-{Dependent} {Friction} {Effects} on {Vibrational} {Infrared}
  {Frequencies} and {Line} {Shapes} of {Liquid} {Water}.
\newblock {\em J. Phys. Chem. B}, 126(7):1579--1589, February 2022.
\newblock Publisher: American Chemical Society.

\bibitem{dieterich_anomalous_2008}
Peter Dieterich, Rainer Klages, Roland Preuss, and Albrecht Schwab.
\newblock Anomalous dynamics of cell migration.
\newblock {\em Proceedings of the National Academy of Sciences},
  105(2):459--463, January 2008.
\newblock Publisher: Proceedings of the National Academy of Sciences.

\bibitem{klimek_2024_cell}
Anton Klimek, Johannes~CJ Heyn, Debasmita Mondal, Sophia Schwartz, Joachim~O
  R{\"a}dler, Prerna Sharma, Stephan Block, and Roland~R Netz.
\newblock Intrinsic cell-to-cell variance from experimental single-cell
  motility data.
\newblock {\em PRX Life}, 3:023015, 2025.

\bibitem{klimek_data-driven_2024}
Anton Klimek, Debasmita Mondal, Stephan Block, Prerna Sharma, and Roland~R
  Netz.
\newblock Data-driven classification of individual cells by their non-markovian
  motion.
\newblock {\em Biophysical Journal}, 123(10):1173--1183, 2024.

\bibitem{guo_entanglement-controlled_2012}
Hongyu Guo, Gilles Bourret, R.~Bruce Lennox, Mark Sutton, James~L. Harden, and
  Robert~L. Leheny.
\newblock Entanglement-{Controlled} {Subdiffusion} of {Nanoparticles} within
  {Concentrated} {Polymer} {Solutions}.
\newblock {\em Phys. Rev. Lett.}, 109(5):055901, August 2012.

\bibitem{abbasi_non-markovian_2023}
Amir Abbasi, Roland~R. Netz, and Ali Naji.
\newblock Non-{Markovian} {Modeling} of {Nonequilibrium} {Fluctuations} and
  {Dissipation} in {Active} {Viscoelastic} {Biomatter}.
\newblock {\em Phys. Rev. Lett.}, 131(22):228202, November 2023.

\bibitem{goychuk_finite-range_2020}
Igor Goychuk and Thorsten Pöschel.
\newblock Finite-range viscoelastic subdiffusion in disordered systems with
  inclusion of inertial effects.
\newblock {\em New J. Phys.}, 22(11):113018, November 2020.
\newblock Publisher: IOP Publishing.

\bibitem{evers_colloids_2013}
F.~Evers, R.~D.~L. Hanes, C.~Zunke, R.~F. Capellmann, J.~Bewerunge,
  C.~Dalle-Ferrier, M.~C. Jenkins, I.~Ladadwa, A.~Heuer, R.~Castañeda-Priego,
  and S.~U. Egelhaaf.
\newblock Colloids in light fields: {Particle} dynamics in random and periodic
  energy landscapes.
\newblock {\em Eur. Phys. J. Spec. Top.}, 222(11):2995--3009, November 2013.

\end{thebibliography}
\end{document}


\maketitle
	

	\section{Fitting-parameter results \& simulation details}\label{sec_sim_details}
	In this section, we tabulate all parameters of the multi-exponential memory Eq.~\eqref{eq_kern_expo} fitted to the extracted friction kernel $\Gamma(t)$ via a least-squares fit.
	Additionally, we explain the simulation setup of the GLE simulations and the Markovian Langevin simulations with coordinate-dependent friction.
	
    In Tab.~\ref{tab_times_prot}, we show all characteristic times of the conformational dynamics of the studied proteins.
    The longest memory time and the first moment of memory times, $\tau_{\rm{mem}} = \int_{0}^{\infty}t\Gamma(t)dt/\int_{0}^{\infty}\Gamma(t)dt$, are on the order of 10 to 100 times smaller than the MFPT, $\tau_{\rm{MFP}}$.
    Even though the memory times are smaller than the respective folding and unfolding times, they strongly influence the dynamics of the MSD, shown in the main text, Fig.~\ref{fig3_protein_msd}, and of the MFPT, shown in Fig.~\ref{fig4_mfpt}.
    In Tabs. \ref{tab_frics_prot} and \ref{tab_frics_ala}, we show the friction amplitudes, $\gamma_i$, from the multi-exponential fits to the friction kernels in the main text, Fig.~\ref{fig2_kernels}

    The simulation time steps $h$ are, for Ala9, $h=100\tau_m$ for coordinate-dependent friction Langevin simulations and $h=10\tau_m$ for GLE simulations and the Langevin simulation with constant friction $\gamma_{\rm{tot}}$.
    All simulations for the other proteins are performed with $h=\tau_m$.
    The initial position and velocity are drawn from the respective Boltzmann distribution, and 1000 independent trajectories of length $10^8h$ are generated to extract average MSD and average MFPT profiles.
    The MSD of each individual trajectory is extracted via time averaging according to
	\begin{equation}
		\label{eq_msd_from_data}
		C_{\rm{MSD}}(ih)=\frac{1}{N - i + 1}\sum_{j=0}^{N-i}(x_{j+i}-x_j)^2\,,
	\end{equation}

    For the integration of all equations of motion, we use a fourth-order Runge-Kutta integrator, where coordinate-dependent quantities, such as the free energy $U(q)$ or the friction profile $\gamma(q)$ in the case of coordinate-dependent friction, are interpolated by cubic splines in order to obtain values at positions between bins.
  
    The values of $c$ and $d$ given in the main text, Tab. 1, result from the inversion of Eq.~\eqref{eq_cd} using the values of $\gamma_i$ and $\tau_i$ from the fits, given in Tabs. \ref{tab_times_prot}, \ref{tab_frics_prot}, \ref{tab_frics_ala}.
    Values of $\alpha_{\rm{sub}}^{\rm{pred}}$ are obtained by averaging Eq.~\eqref{eq_inverse_cd} over $1<i<4$, and errors are estimated via the standard deviation.
    The ratios of consecutive memory times and consecutive friction coefficients are shown in Tab.~\ref{tab_ratios_all} to exhibit similar values across all scales and even across different proteins.
    These constant ratios lead to an even spacing between individual memory components on the log-log scale of $t$ and $\Gamma (t)$.
    This is depicted in Fig.~\ref{fig2_kernels}A of the main text, where the violet line corresponds to $\frac{\gamma_3}{\tau_3}e^{-t/\tau_3}$, the cyan line corresponds to $\frac{\gamma_3}{\tau_3}e^{-t/\tau_3} + \frac{\gamma_2}{\tau_2}e^{-t/\tau_2}$ and the red line corresponds to the complete sum $\sum\limits_{i=1}^{3}\frac{\gamma_i}{\tau_i}e^{-t/\tau_i}$.
    The Ala9 data involve a shortest and longest memory component that do not affect the subdiffusive regime in the MSD and thus are not used to predict the subdiffusive scaling.
    For all other proteins, all exponential memory components lie within the subdiffusive regime of the MSD and are used to compute $c$ and $d$ and predict the subdiffusive scaling $\alpha$ via Eq.~\eqref{eq_inverse_cd} in the main text.
	
	\begin{table}
		\centering
		\begin{tabular}{ |c|c|c|c|c|c|c|c|c|c|c| }
			\hline
			&  &  &  &  &  &  &  &  &  &  \vspace*{-3mm}\\
			Protein & $\tau_m/\rm{ps}$ & $\tau_D/\rm{\mu s}$ & $\tau_1/\rm{ns}$ & $\tau_2/\rm{ns}$ & $\tau_3/\rm{ns}$ & $\tau_4/\rm{ns}$ & $\tau_5/\rm{ns}$ & $\tau_{\rm{mem}}/\rm{ns}$ & $\tau_{\rm{MFP}}^{\rm{fol}}/\mu\rm{s}$ & $\tau_{\rm{MFP}}^{\rm{unf}}/\mu\rm{s}$ \\
			\hline
			$\lambda$-repressor & 4.9 & $18.8$ & $2.2$  & $28$ & $260$ &  &  & 198 & 24 & 8 \\ 
			\hline
			$\alpha_3$D & 1.5 & 38.7 & 4.9 & 63 & 4300 &  &  & 3760 & 25 & 24 \\ 
			\hline 
			Protein-G & 3.2 & 15.5 & 2.8 & 43 & 410 &  &  & 343 & 47 & 33 \\ 
			\hline
			Ala9 & $9\times 10^{-5}$ & 0.064 & $7\times 10^{-6}$ & 0.005 & 0.040 & 0.399 & 4.970 & 1 & 0.06 & 0.02 \\
			\hline
		\end{tabular}
		\caption{Inertial time $\tau_m$, diffusion time $\tau_D$, and fitted memory times $\tau_i$ in the multi-exponential friction kernel $\Gamma(t) = \sum_{i=1}^{n}\frac{\gamma_i}{\tau_i} e^{-t/\tau_i}$ for different proteins. The first moment of fitted memory times is given by $\tau_{\rm{mem}} = \int_{0}^{\infty}t\Gamma(t)dt/\int_{0}^{\infty}\Gamma(t)dt$. The MFPTs from the folded state free energy minimum to the unfolded state minimum $\tau_{\rm{MFP}}^{\rm{fol}}$ and vice versa $\tau_{\rm{MFP}}^{\rm{unf}}$ are extracted from MD simulations.}
		\label{tab_times_prot}
	\end{table}

\begin{table}
	\centering
	\begin{tabular}{ |c|c|c|c|c|c|c|c|c| }
		\hline
		&  &  &  &  &  &  &  &  \vspace*{-3mm}\\
		Protein & $\tau_5/\tau_4$ & $\tau_4/\tau_3$ & $\tau_3/\tau_2$ & $\tau_2/\tau_1$ & $\gamma_5/\gamma_4$ & $\gamma_4/\gamma_3$ & $\gamma_3/\gamma_2$ & $\gamma_2/\gamma_1$ \\
		\hline
		$\lambda$-repressor &  &  & 9.3  & 12.7 &  &  & 3.7 & 2.5 \\
		\hline
		$\alpha_3$D &  &  & 68.3 & 12.8 &  &  & 8.2 & 4.4 \\ 
		\hline 
		Protein-G &  &  & 9.5 & 15.4 &  &  & 5.1 & 5.2 \\ 
		\hline
		Ala9 & 12.5 & 10.0 & 8.0 & 714.2 & 0.2 & 5.7 & 3.5 & 5.5 \\
		\hline
	\end{tabular}
	\caption{Ratios of memory times and of friction coefficients.}
	\label{tab_ratios_all}
\end{table}

\begin{table}
	\centering
	\begin{tabular}{ |c|c|c|c|c| }
		\hline
		&  &  &  &  \vspace*{-3mm} \\
		Protein & $\gamma_1$ & $\gamma_2$ & $\gamma_3$ & $\gamma_{\rm{tot}}$ \\
		\hline
		$\lambda$-repressor & 60 & 148 & 542  & 750  \\ 
		\hline
		$\alpha_3$D & 54 & 239 & 1960 & 2253 \\ 
		\hline 
		Protein-G & 22 & 113 & 588 & 723  \\ 
		\hline

	\end{tabular}
	\caption{Friction components $\gamma_i$ of multi-exponential fit $\Gamma(t) = \sum_{i=1}^{n}\frac{\gamma_i}{\tau_i} e^{-t/\tau_i}$ and total friction $\gamma_{\rm{tot}}$ for different proteins in units $\rm{u}\,\rm{nm}^2/s$.}
	\label{tab_frics_prot}
\end{table}

\begin{table}
 	\centering
 	\begin{tabular}{ |c|c|c|c|c|c|c| }
 		\hline
 		&  &  &  &  &  &  \vspace*{-3mm} \\
 		 & $\gamma_1$ & $\gamma_2$ & $\gamma_3$ & $\gamma_4$ & $\gamma_5$ & $\gamma_{\rm{tot}}$ \\
 		\hline
		Ala9 & 2.2 & 12 & 42 & 240 & 57 & 350 \\
		\hline
	\end{tabular}
	\caption{Friction components $\gamma_i$ of multi-exponential fit $\Gamma(t) = \sum_{i=1}^{n}\frac{\gamma_i}{\tau_i} e^{-t/\tau_i}$ and total friction $\gamma_{\rm{tot}}$ for Ala9 in units $\rm{u}/\rm{ns}$.}
	\label{tab_frics_ala}
\end{table}

	\section{MFPT and friction profiles from MD data}
	The MFPT is defined as the mean time to reach a final position $q_F$ for the first time when starting from a position $q_S$. For the Langevin Eq.~\eqref{eq_pos_dep_motion} with coordinate-dependent friction, the MFPT for $q_S < q_F$ is given by
    \begin{equation}
        \label{eq_mfpt_OD_theo}
        \tau_{\text {MFP}}\left(q_S, q_F\right)=\beta \int_{q_S}^{q_F} \mathrm{~d} q e^{\beta U(q)} \gamma(q) \int_{q_{\min }}^q \mathrm{~d} q^{\prime} e^{-\beta U\left(q^{\prime}\right)}
    \end{equation}
    and for $q_S>q_F$,
    \begin{equation}
        \tau_{\rm{MFP}}\left(q_S, q_F\right)=\beta \int_{q_F}^{q_S} \mathrm{~d} q e^{\beta U(q)} \gamma(q) \int_q^{q_{\max }} \mathrm{d} q^{\prime} e^{-\beta U\left(q^{\prime}\right)}\, .
    \end{equation}
	
\begin{figure*}
		\vspace{-2cm}
		\includegraphics{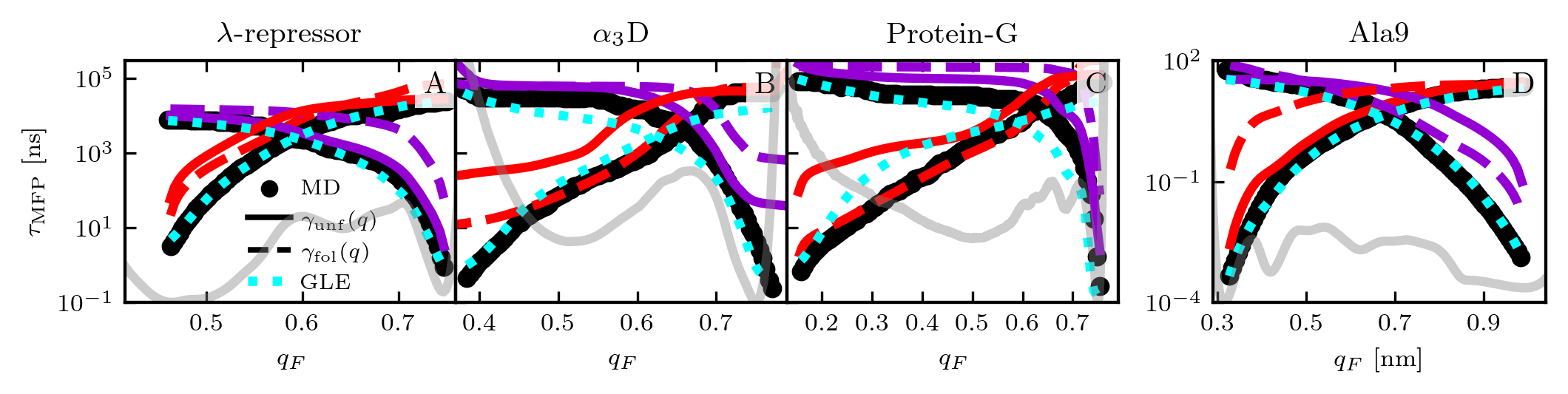}
		\caption{The MFPT extracted from MD simulations (black dots) as a function of $q_F$, is compared to the prediction from GLE simulation (cyan dotted lines) and the prediction from the Langevin equation with coordinate-dependent friction from folding (dashed lines) and unfolding (solid lines).
        Starting points for the folding profiles are: for $\lambda$-repressor, $q_S=0.46$; for $\alpha_3$D, $q_S=0.38$; for protein-G, $q_S=0.15$; and for Ala9, $q_S=0.99,\rm{nm}$.
        Starting points for the unfolding profiles are: for $\lambda$-repressor, $q_S=0.75$; for $\alpha_3$D, $q_S=0.78$; for protein-G, $q_S=0.76$; and for Ala9, $q_S=0.32,\rm{nm}$.
        The gray line in the background shows the shape of the free energy $U(q)$.}
		\label{fig4_mfpt}
	\end{figure*}

\begin{figure*}
		\vspace{-2mm}
		\includegraphics{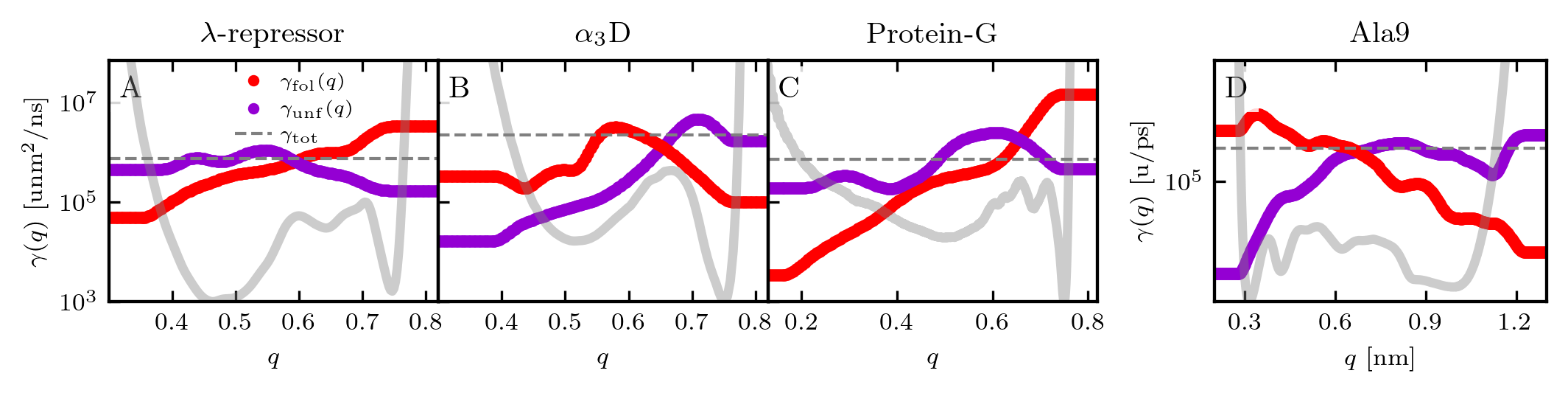}
		\caption{Coordinate-dependent friction profiles extracted from MFPT profiles of the MD data. The profiles extracted in the folding direction are shown in red; profiles from the unfolding direction, in violet. The dashed line denotes the total friction $\gamma_{\rm{tot}}$, and the gray line shows the shape of the free energy $U(q)$.}
		\label{fig5_gammaq}
	\end{figure*}

    The derivative of Eq.~\eqref{eq_mfpt_OD_theo} with respect to $q_F$ yields the friction profile $\gamma\left(q_F\right)$ as \cite{hinczewski_how_2010,ayaz_non-markovian_2021}
    \begin{equation}\label{eq_gammaq_extraction}
        \begin{gathered}
            \gamma_{\text {unf}}\left(q_F\right)=k_B T \frac{e^{-\beta U\left(q_F\right)}}{Z_1} \frac{\partial \tau_{\mathrm{MFP}}}{\partial q_F} \text { for } q_S<q_F, \\
            \gamma_{\mathrm{fol}}\left(q_F\right)=-k_B T \frac{e^{-\beta U\left(q_F\right)}}{Z_2} \frac{\partial \tau_{\mathrm{MFP}}}{\partial q_F} \text { for } q_S>q_F\,,
        \end{gathered}
    \end{equation}
    where $Z_1=\int_{q_{\min }}^{q_F} \mathrm{~d} q e^{-\beta U(q)}$ and $Z_2=\int_{q_F}^{q_{\max }} \mathrm{d} q e^{-\beta U(q)}$.
    Eqs.~\eqref{eq_gammaq_extraction} are then used to compute coordinate-dependent friction profiles from the MFPT profiles shown in Fig.~\ref{fig4_mfpt}.
    The resulting folding and unfolding friction profiles are shown in Fig.~\ref{fig5_gammaq}.
	
    The MD-simulation MFPT profiles in Fig.~\ref{fig4_mfpt}A,D for $\lambda$-repressor and Ala9 are well reproduced by GLE simulations (cyan dotted lines).
    For $\alpha_3D$ and protein-G, the MD-simulation MFPT and GLE-simulation MFPT show good agreement for small distances between the start and final positions ($q_S$ and $q_F$) and long distances between $q_S$ and $q_F$, while there are some deviations for intermediate distances.
    In contrast, the Markovian Langevin equation with coordinate-dependent friction cannot describe the MFPT profiles consistently, since neither $\gamma_{\rm{fol}}(q)$ alone nor $\gamma_{\rm{unf}}(q)$ alone can describe both the MFPT profile in the folding and unfolding directions.
    If the coordinate-dependent friction $\gamma(q)$ is extracted from a single MFPT folding profile, it exactly reproduces the input MFPT profile by construction in Eq.~\eqref{eq_gammaq_extraction}; however, it does not reproduce the MFPT in the opposite direction \cite{ayaz_non-markovian_2021}.
    In order to obtain a smooth coordinate-dependent friction profile, we average over several MFPT profiles with different starting points $q_S$, as explained in Sec.~\ref{sec_gammaq_from_mfpt}.
    This leads to a close agreement of the coordinate-dependent prediction of the MFPT with the MD input data; dashed lines are close to folding MFPT profiles, and solid lines are close to unfolding MFPT profiles in Fig.~\ref{fig4_mfpt}.
    Note that for the $\rm{HB}_4$ RC of Ala9, the folding occurs with decreasing $q$, whereas for the fraction of native contacts RC, folding is described by increasing $q$.
    Coordinate-dependent friction profiles from folding MFPT profiles lead to poor agreement for unfolding MFPT profiles and vice versa, as shown in Fig.~\ref{fig4_mfpt}.
    Thus, the theory of coordinate-dependent friction does not consistently describe the dynamics, as seen in the main text for the MSD and as further underscored by the friction profiles themselves, which largely deviate when extracted from the folding or unfolding direction, as seen in Fig.~\ref{fig5_gammaq}.
    In contrast, the GLE captures the dynamics of folding and unfolding MFPT in Fig.~\ref{fig4_mfpt} and the MSD, as shown in the main text Fig.~\ref{fig3_protein_msd}.
    This suggests that the observed position dependency of the extracted friction profiles is simply an artifact of assuming a Markovian coordinate-dependent friction in a system that is actually well described by the GLE with position-independent memory kernel.

	\section{Extracting friction profiles from MFPT profiles}
	\label{sec_gammaq_from_mfpt}
    
\begin{figure*}
		\centering
		\includegraphics{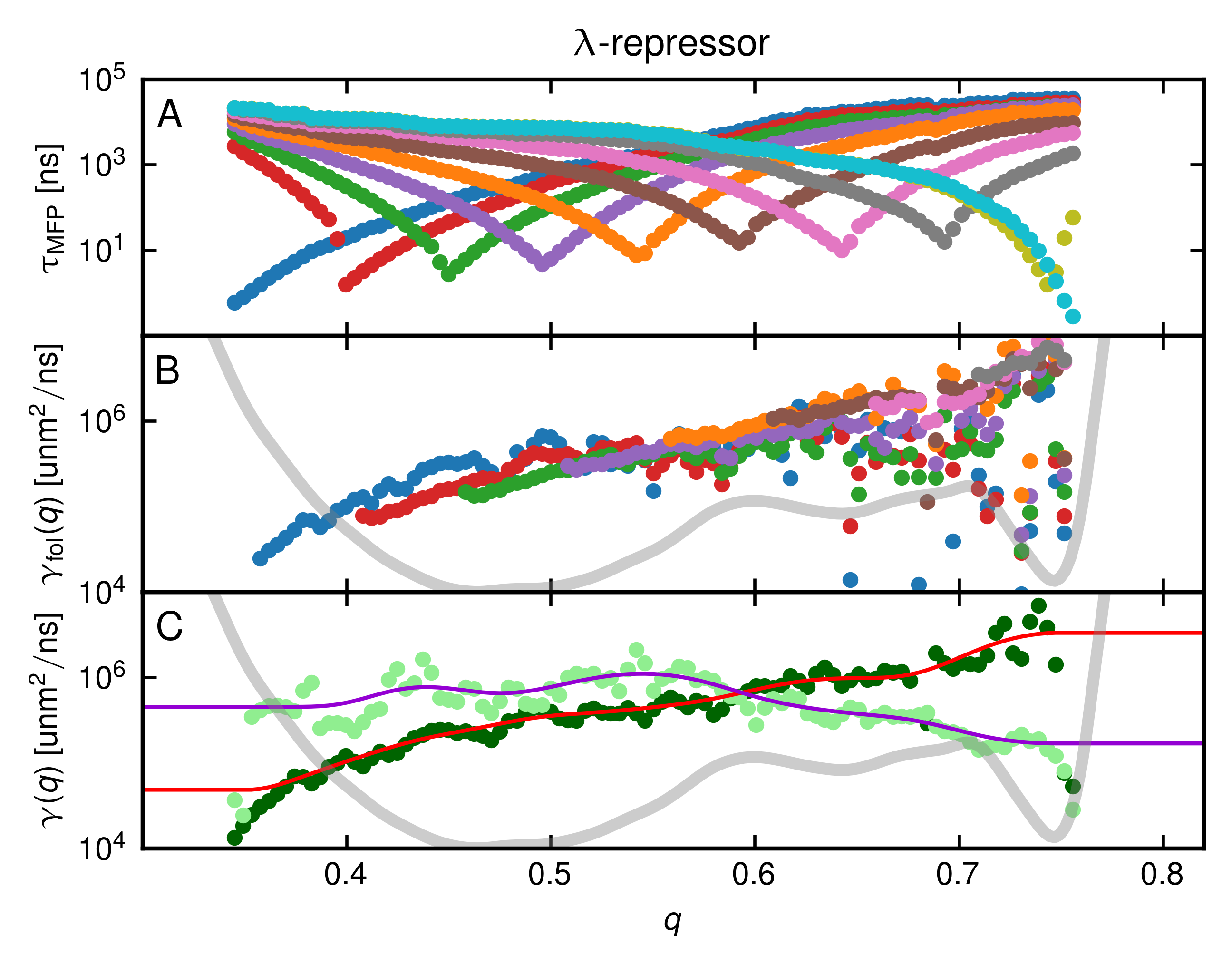}
		\caption{(A) MFPT profiles from MD simulations with different starting points $q_S$ are shown in different colors.
        Starting points $q_S$ are located at the minimum of the respective profile.
        The parts of the MFPT curves for which $\tau_{\rm{MFP}}$ increases for increasing $q$ is used to compute a folding friction profile in the respective $q$ regime.
        Parts where $\tau_{\rm{MFP}}$ increases for decreasing $q$ are used to compute unfolding friction profiles in the respective $q$ regimes.
        (B) All folding friction profiles extracted from the folding MFPT profiles in (A) are shown in the corresponding color.
        (C) The average over friction profiles with different starting points $q_S$ is shown in the lower panel for the folding direction in dark green (average over profiles in (B)) and unfolding in light green.
        The red and violet lines represent the smoothed friction profiles that are extrapolated as constant on both sides, which are used for simulations of Eq.~\eqref{eq_pos_dep_motion}.}
		\label{fig_explain_fric}
	\end{figure*}

    Friction profiles are extracted from MFPT profiles according to Eq.~\eqref{eq_gammaq_extraction} \cite{hinczewski_how_2010}.
    In order to decrease the noise in coordinate-dependent friction profiles, we extract MFPT profiles for different starting points $q_S$, as shown in Fig.~\ref{fig_explain_fric}A.
    This yields several $\gamma(q)$ curves in overlapping position intervals, as shown in Fig.~\ref{fig_explain_fric}B for the folding direction. All the friction profiles in one direction, as shown in Fig.~\ref{fig_explain_fric}B, are averaged to obtain the coordinate-dependent friction profiles shown in Fig.~\ref{fig_explain_fric}C in the respective direction (dark green for folding and light green for unfolding).
    The MD simulations only rarely sample positions of high free energy, which increases the noise of the coordinate-dependent friction in regions with high free energy, especially for high and low values of $q$.
    Certain RC values are not covered at all in the MD data because of the finite simulation length.
    In order to regularize the behavior of our Langevin simulations at these points, we need to define the coordinate-dependent quantities $\gamma(q)$ and $U(q)$ for all possible RC values.
    In Langevin simulations with coordinate-dependent friction, governed by Eq.~\eqref{eq_pos_dep_motion}, we augment the friction profiles as constant on both sides, where MD simulations cannot provide the necessary MFPT information.
    The free energy profile $U(q)$ is augmented on both sides by the continuation of the cubic spline interpolation of the last available RC values.
    Finally, we smooth the friction profiles with a Gaussian window function, which yields the red and violet lines in Fig.~\ref{fig_explain_fric}C that correspond to the lines shown in Fig.~\ref{fig5_gammaq}A. \\

	\section{Coordinate-dependent friction with inertial effects does not improve agreement with MD data}
    
	An exact expression for the MFPT is only available in the overdamped scenario described by Eq.~\eqref{eq_pos_dep_motion}.
    However, once the coordinate-dependent friction $\gamma(q)$ is extracted from the overdamped Eq.~\eqref{eq_gammaq_extraction}, we can add inertial effects to the equation of motion to test the influence of inertial effects in the presence of coordinate-dependent friction.
    The Langevin equation including inertial effects and coordinate-dependent friction $\gamma(q)$ reads
	\begin{equation}
		\label{eq_le_posdep_UD}
		m \ddot{q}(t) = -\nabla U(q) - \gamma(q) \dot{q}(t)+ \sqrt{k_B T \gamma(q)} \xi(t)\,.
	\end{equation}
	We simulate Eq.~\eqref{eq_le_posdep_UD} by using the same mass $m=k_BT/\langle\dot{q}^2\rangle$ as for the GLE simulations and the friction profiles shown in Fig.~\ref{fig5_gammaq}, extracted from the overdamped theory Eq.~\eqref{eq_gammaq_extraction}.

\begin{figure*}
		\centering
		\includegraphics{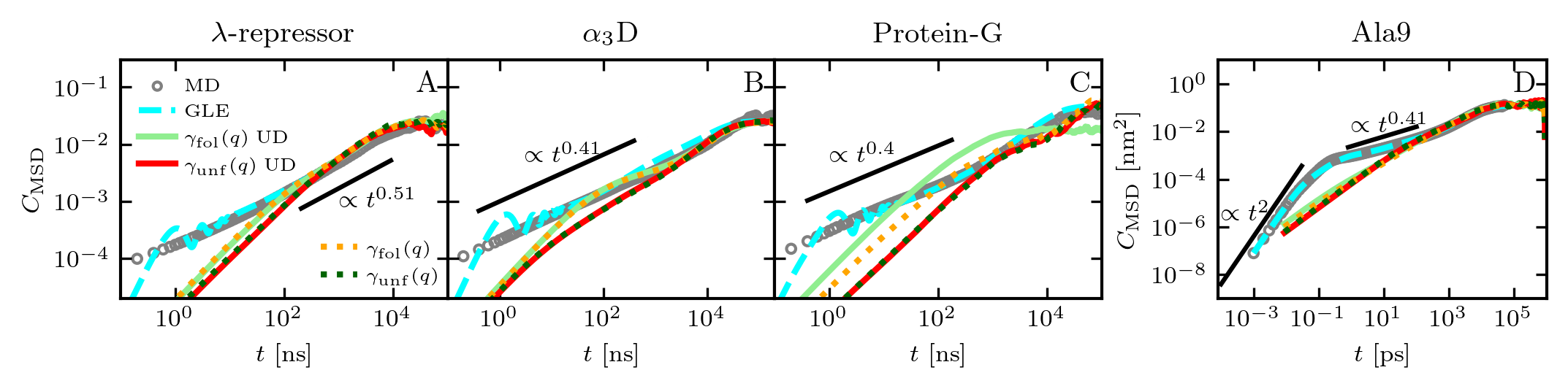}
		\caption{MSD of different proteins.
        Data extracted from MD simulations are shown as gray circles, where the cyan dashed lines represent the result from simulating the GLE in Eq.~\eqref{eq_gle_wom} in the main text using the respective fitted kernel $\Gamma(t)$ of Figs.~\ref{fig2_kernels}A-D in the respective potential $U(x)$ of Figs.~\ref{fig1}B-D.
        The dotted lines represent Markovian Langevin simulation results with the respective extracted coordinate-dependent friction profile shown in Fig.~\ref{fig5_gammaq} in the respective potential landscape of Figs.~\ref{fig1}B-D, where light green and red lines represent underdamped dynamics, including inertial effects according to Eq.~\eqref{eq_le_posdep_UD}, and dark green and orange lines represent the overdamped dynamics of Eq.~\eqref{eq_pos_dep_motion}.}
		\label{fig_msd_posdep_ud}
	\end{figure*}

    The MSD of simulations of Eq.~\eqref{eq_le_posdep_UD} shows very similar behavior to the MSD of the overdamped version Eq.~\eqref{eq_pos_dep_motion}, as seen in Fig.~\ref{fig_msd_posdep_ud}.
    This showcases the overdamped nature of protein folding dynamics, as demonstrated in the main text already by $\tau_m \ll \tau_D$.
    Furthermore, the addition of inertial effects in the framework of the Langevin equation with coordinate-dependent friction still cannot reproduce the correct subdiffusive scaling exponents.
    Only in the long-time limit do Markovian theories capture the correct behavior, in which case the dynamics are effectively described by a constant friction and the confinement of the external potential.
	
\begin{figure*}
		\centering
		\includegraphics{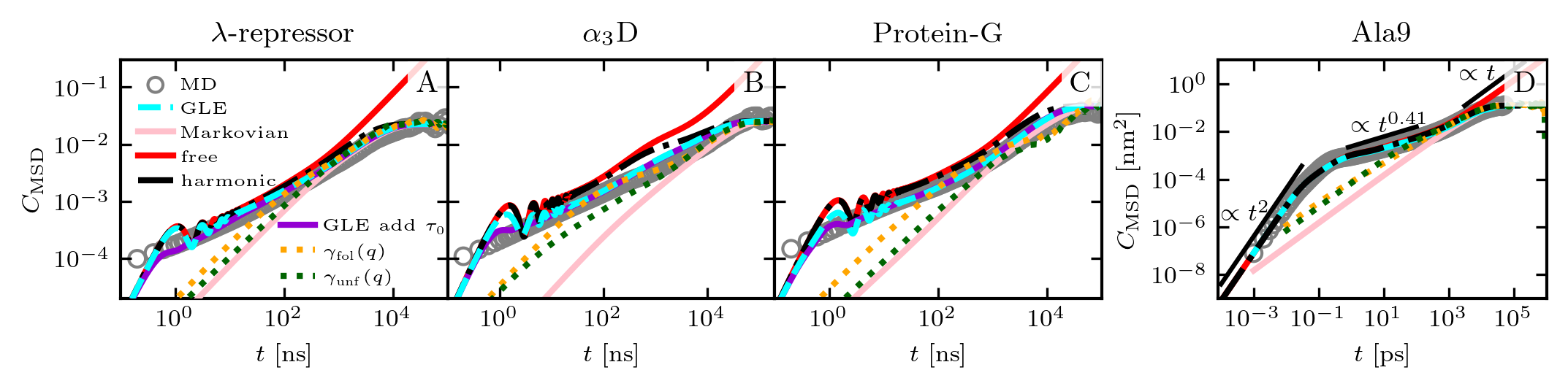}
		\caption{MSD of different proteins.
        Data extracted from MD simulations are shown as gray circles, where the cyan dashed lines represent the GLE simulation results using the respective fitted kernel $\Gamma(t)$ of Figs.~\ref{fig2_kernels}A-D in the respective potential landscape, Figs.~\ref{fig1}B-D.
        The violet line results from simulating a kernel with an additional memory component on a shorter time scale according to Eq.~\eqref{eq_kern_expo}, with values of $c$ and $d$ in Eq.~\eqref{eq_cd} taken from Tab.~\ref{tab_alpha_protein}.
        The red line is the analytical prediction without an external potential; the black dash-dotted line is the analytical result in a harmonic potential with the same positional variance as the free energy \cite{klimek2024_theo}. The dotted lines result from the simulation of Eq.~\eqref{eq_pos_dep_motion} with coordinate-dependent friction profiles (shown in the SI) extracted from the unfolding MFPT profile and folding MFPT profile (shown in the SI as well) according to Eq.~\eqref{eq_gammaq_extraction} in the respective color.
        The pink lines stem from simulations of the Langevin equation in the Markovian limit using the constant total friction $\gamma_{\rm{tot}}$.
        All simulation results are averaged over 1,000 independent trajectories, each of length $10^8$ steps; further simulation details are given in Sec. \ref{sec_sim_details}.}
		\label{fig3_protein_msd_all}
	\end{figure*}

    In Fig.~\ref{fig3_protein_msd_all}, we compile the MSD of all models discussed in the main text.
    As explained in the main text, the GLE simulations agree nicely with the MD data (gray circles), and the agreement is improved by adding a friction memory component on a shorter time scale according to the mean ratios of fitted memory time scales, $c$, and friction amplitudes, $d$, in Tab. 1 of the main text.
    The Markovian models only capture the long-time behavior accurately and do not describe the subdiffusive scaling.
    The red lines show the analytical result \cite{klimek2024_theo} of the GLE with $U(q)=0$, and the black lines represent the case for $U(q)=Kq^2/2$, with $K$ chosen to reproduce the long-time MSD plateau value, using the variance of the position distribution $p(x)$ in Eq.~\eqref{eq_boltzmann_inverted} for the respective extracted free energy $U(q)$.

	\section{GLE captures time-dependent exponent $\alpha(t)$}

\begin{figure*}
		\centering
		\includegraphics{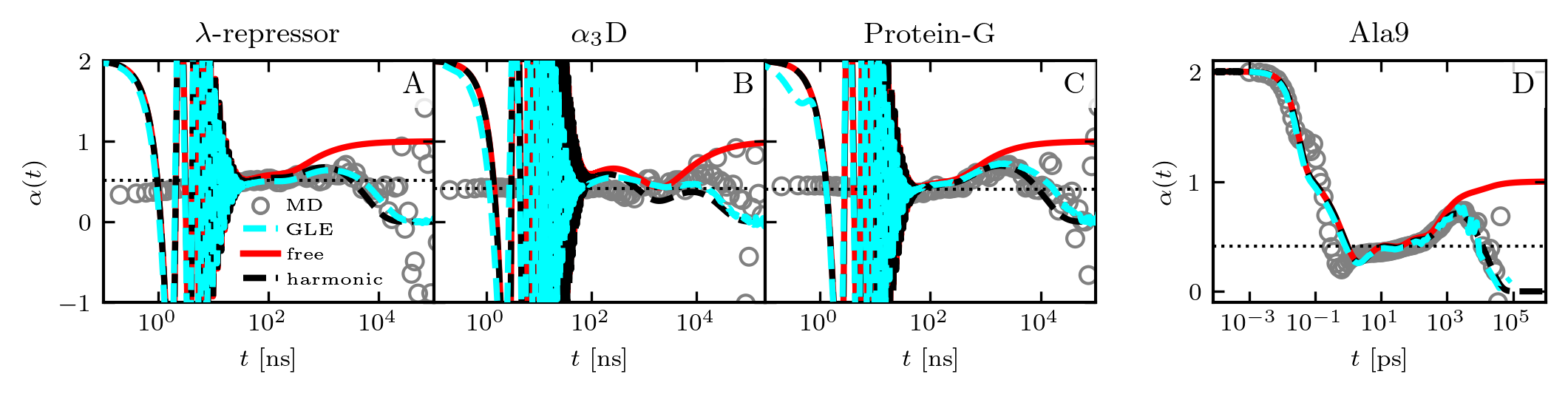}
		\caption{Comparison of $\alpha(t)$ extracted via Eq.~\eqref{eq_alpha} from MD data as gray circles, GLE simulations with fitted kernel $\Gamma(t)$ and extracted potential $U(q)$ as cyan lines, analytical prediction for $\Gamma(t)$ with $U(q)=0$ in red and analytic prediction for $\Gamma(t)$ with a harmonic potential $U(q)=Kq^2/2$ in black \cite{klimek2024_theo}.
        The horizontal dotted lines show the respective value of $\alpha_{\rm{sub}}^{\rm{data}}$.}
		\label{fig_protein_alpha_all}
	\end{figure*}

    Since the GLE predicts the MSD of the proteins well, as shown in the main text, it consequently captures the time-dependent exponent $\alpha (t)$ as well, which we demonstrate in Fig.~\ref{fig_protein_alpha_all}.
    Only the oscillations at short times in $\alpha(t)$ of the GLE simulations deviate from the behavior of the MD data.
    These oscillations are due to the missing information about shorter time scales because of the finite time resolution of the MD data and can be reduced by adding shorter memory times, as explained in the main text.
    The analytic predictions using the fitted friction kernel $\Gamma(t)$ with $U(q)=0$ (red lines) describe $\alpha(t)$ well up to roughly $1\unit{ns}$ for Ala9 and up to $1\unit{\mu s}$ for all other proteins.
    The analytic prediction using $\Gamma(t)$ and a harmonic potential (black dashed line) describes the exponent $\alpha (t)$ well for all times, as it captures the long-time confinement effect leading to a plateau in the MSD and thus, a value of $\alpha=0$ for long times.
    This indicates that $\alpha (t)$ is dictated mainly by friction memory effects, and the main influence of the free energy landscape is the confinement for long times.
    As discussed in the main text, higher free energy barriers can introduce additional subdiffusive regimes; however, for the studied fast-folding proteins, the scaling behavior is completely governed by the memory.
    For short times, before coupling to the local shape of deep free energy wells, overdamped dynamics are completely governed by memory, as explained in \cite{klimek2024_theo}.
    Since the dynamics of proteins are usually overdamped, as also shown in the main text for the studied proteins in Tab. 1, the short-time dynamics have to be governed by memory effects, and the long-time dynamics are still strongly influenced by the memory.
    Only the knowledge of both the free energy landscape $U(q)$ and friction memory $\Gamma(t)$ allows for a correct description of protein folding dynamics.

	\section{Addition of longer memory time contributions reduces agreement with MD data}
    
\begin{figure*}
		\centering
		\includegraphics{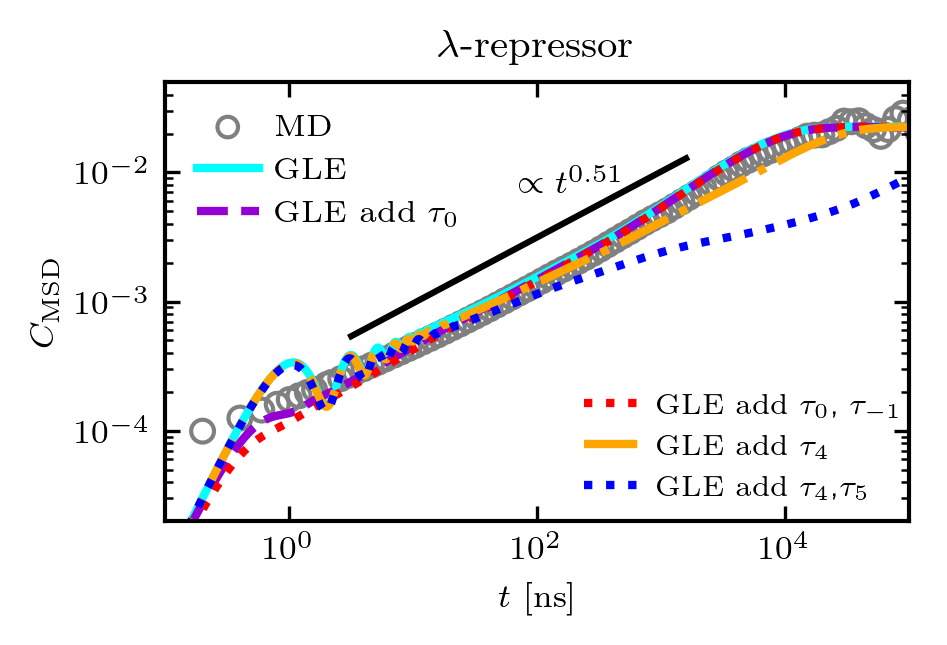}
		\caption{MSD for the fraction of native contacts from MD simulations of $\lambda$-repressor \cite{lindorff-larsen_how_2011} as gray circles.
        The cyan line shows the MSD from a simulation of the GLE with the fitted memory kernel Eq.~\eqref{eq_kern_expo} shown in Fig.~\ref{fig2_kernels}a and the free energy profile shown in Fig.~\ref{fig1}C.
        All other lines show the MSD from simulations with additional time scales according to the observed hierarchical pattern of memory contributions according to Eq.~\eqref{eq_kern_expo} with values of $c$ and $d$ in Eq.~\eqref{eq_cd} taken from Table \ref{tab_alpha_protein}.
		}
		\label{fig_SI_add_lambda}
	\end{figure*}

    Adding memory contributions according to the observed hierarchical pattern in the memory times and amplitudes measured in the the data leads to better agreement with the MD data for the fast-folding proteins when short time scales are added, as shown in Figs.~\ref{fig3_protein_msd_all} by the violet and red lines.
    In contrast, adding memory contributions on longer time scales than the largest fitted memory time, $\tau_3$ in the case of $\lambda$-repressor, leads to worse agreement with the MSD of the MD data.
    The addition of one memory contribution on a longer time scale leads to an MSD (orange line  in Fig.~\ref{fig_SI_add_lambda}), which shows a similar deviation from the MD data as the GLE prediction using the tri-exponential fit (cyan line).
    However, the addition of this memory contribution on a longer time scale increases the total friction approximately $d$-fold, with $d=2.7$ for $\lambda$-repressor.
    We can estimate the order of magnitude of the total friction by averaging over the coordinate-dependent friction profiles as $\gamma_{\textrm{tot}}^{\textrm{est}} = \int_{q_{\rm{min}}}^{q_{\rm{max}}} dq \gamma(q) / \int_{q_{\rm{min}}}^{q_{\rm{max}}} dq$, where $q_{\rm{min}}$ and $q_{\rm{max}}$ are the values between which the position dependent friction profile was extracted.
    For $\lambda$-repressor, $\gamma_{\textrm{tot}}^{\textrm{est}} = 7.4\times10^{5}\unit{unm^2/ns}$ (averaging over folding and unfolding direction) and the total friction predicted from the integral of the kernel fit is $\gamma_{\rm{tot}}=7.5\times 10\unit{unm^2/ns}$.
    This means, the addition of memory components on a longer time scale that include high friction amplitudes, leads to a total friction that is much higher than we would expect from the data.
    Moreover, adding two contributions on longer time scales, shown by the dark-blue line in Fig.~\ref{fig_SI_add_lambda}, leads to strong deviations from the MD MSD, especially for long times.
    This indicates that there is a largest memory time-scale in the system within the resolved temporal regime.

    The MSD with the additional shorter time scales in the friction kernel (purple and red lines in Fig.~\ref{fig_SI_add_lambda}) shows reduced oscillations compared to the MSD of the other GLE simulations.
    There are slight deviations in the MSDs with additional shorter memory contributions on the scale of roughly $1\unit{ns}$ compared to the MD data.
    These originate most likely from an underestimation of the slope of the ballistic MSD regime, i.e of the mean squared velocity $\langle \dot{q}^2 \rangle$, due to the low temporal resolution; see SI of \cite{klimek_2024_cell}.
    The MD data show no ballistic regime in the MSD.
    This suggests that the time step $\Delta$ is larger than the persistence time and means that the estimation of $\langle \dot{q}^2 \rangle$ involves the distance traveled not only ballistically but also subdiffusively during one time step.
    This yields an underestimate of the value for $\langle \dot{q}^2 \rangle$, which results in the simulation MSDs showing a ballistic regime up to the point where it crosses the MSD of the MD data. 
    Therefore, our GLE model cannot capture the value at the first time step of the MD data $C_{\rm{MSD}}(\Delta)$ for $\alpha_3$D, $\lambda$-repressor and protein-G, given the time resolution of the MD data.
    In contrast, the short-time behavior of the MD MSD of Ala9 is perfectly captured by the GLE description, since the MD time step is much smaller than the end-time of the ballistic regime.
	
	